\documentclass[prl,amsmath,amssymb, notitlepage, twocolumn,superscriptaddress,longbibliography]{revtex4-1}
\usepackage{braket}
\usepackage{amsmath,amsfonts, amssymb, amsthm, dsfont}
\usepackage{yfonts}
\usepackage{bm}
\usepackage{mathrsfs}
\usepackage{graphicx}
\usepackage{verbatim}
\usepackage{tikz}
\usetikzlibrary{calc}
\usepackage{multirow}

\newcommand{\ra}{\rangle}
\newcommand{\comments}[1]{}

\newcommand{\mc}[1]{\mathcal{#1}}

\newcommand{\beq}{\begin{eqnarray}}
\newcommand{\eeq}{\end{eqnarray}}

\newcommand{\Tr}{{\rm Tr}}

\newcommand{\tvect}[2]{\begin{pmatrix} #1\\#2\end{pmatrix}}
\newcommand{\norm}[1]{\left\lVert #1 \right\rVert}
\theoremstyle{definition}

\usepackage{color}
\definecolor{darkblue}{rgb}{0.,0.,0.4}
\definecolor{darkred}{rgb}{0.5,0.,0.}
\usepackage[pdftex,colorlinks=true,linkcolor=darkblue,citecolor=blue,urlcolor=darkred]{hyperref}

\newcommand{\refeq}[1]{Eq.~\eqref{#1}}
\newcommand{\reffig}[1]{Fig.~\ref{#1}}
\newcommand{\beqa}{\begin{equation}\begin{aligned}}
\newcommand{\eeqa}{\end{aligned}\end{equation}}

\newcommand{\dirac}[3]{\bra{#1}#2\ket{#3}}
\graphicspath{{pics/}{}}

\usetikzlibrary{decorations.pathreplacing,decorations.markings}
\tikzset{middlearrow/.style={
		decoration={markings,
			mark= at position 0.55 with {\arrow{#1}} ,
		},
		postaction={decorate}
	}
}

\begin{document}
	
\title{Hamiltonian Tomography via Quantum Quench}
\author{Zhi Li}\affiliation{\PITT}\affiliation{\PQI}\affiliation{\PI}
\author{Liujun Zou}\affiliation{\PI}
\author{Timothy H. Hsieh}\affiliation{\PI}

\newcommand*{\PI}{Perimeter Institute for Theoretical Physics, Waterloo, Ontario N2L 2Y5, Canada}
\newcommand*{\PITT}{Department of Physics and Astronomy, University of Pittsburgh, Pittsburgh, Pennsylvania 15260, United States}
\newcommand*{\PQI}{Pittsburgh Quantum Institute, Pittsburgh, Pennsylvania 15260, United States}

% \date{\today}
	
\begin{abstract}
We show that it is possible to uniquely reconstruct a generic many-body local Hamiltonian from a single pair of generic initial and final states related by evolving with the Hamiltonian for any time interval. We then propose a practical version of the protocol involving multiple pairs of such initial and final states. Using the eigenstate thermalization hypothesis, we provide bounds on the protocol's performance and stability against errors from measurements and in the ansatz of the Hamiltonian. The protocol is efficient (requiring experimental resources scaling polynomially with system size in general and constant with system size given translation symmetry) and thus enables analog and digital quantum simulators to verify implementation of a putative Hamiltonian.

\end{abstract}
	
\maketitle

% \tableofcontents

The advent of quantum many-body simulators  \cite{Blatt12, Zhang2017_DPT,Islam583, Devoret13, Gambetta2017, Greiner2002, BlochColdAtoms, Bernien2017, esslinger, Gross995, greiner, roushan} has enabled the exploration of complex quantum dynamics beyond the capabilities of classical computers.  Given this significant potential, it is vital to determine accurately the Hamiltonian actually being realized by a simulator. However, Hamiltonian tomography for a generic many-body system is challenging, precisely due to the fact that the complexity of many-body dynamics makes benchmarking with a classical computer difficult. Thus far, most progress has been made in systems with either a special Hamiltonian or a small size  \cite{Wchirmer2003, Devitt2005, Cole2005,  Cole2006,  Franco2009, Burgarth2009, Burgarth2009b, Burgarth2011, Silva2011, Senko2014, Zhang2014, Wiebe2014, Wang2015, Ma2017, hauke1, hauke2}.

\begin{figure}
    \centering
    \includegraphics[width=1.01\linewidth]{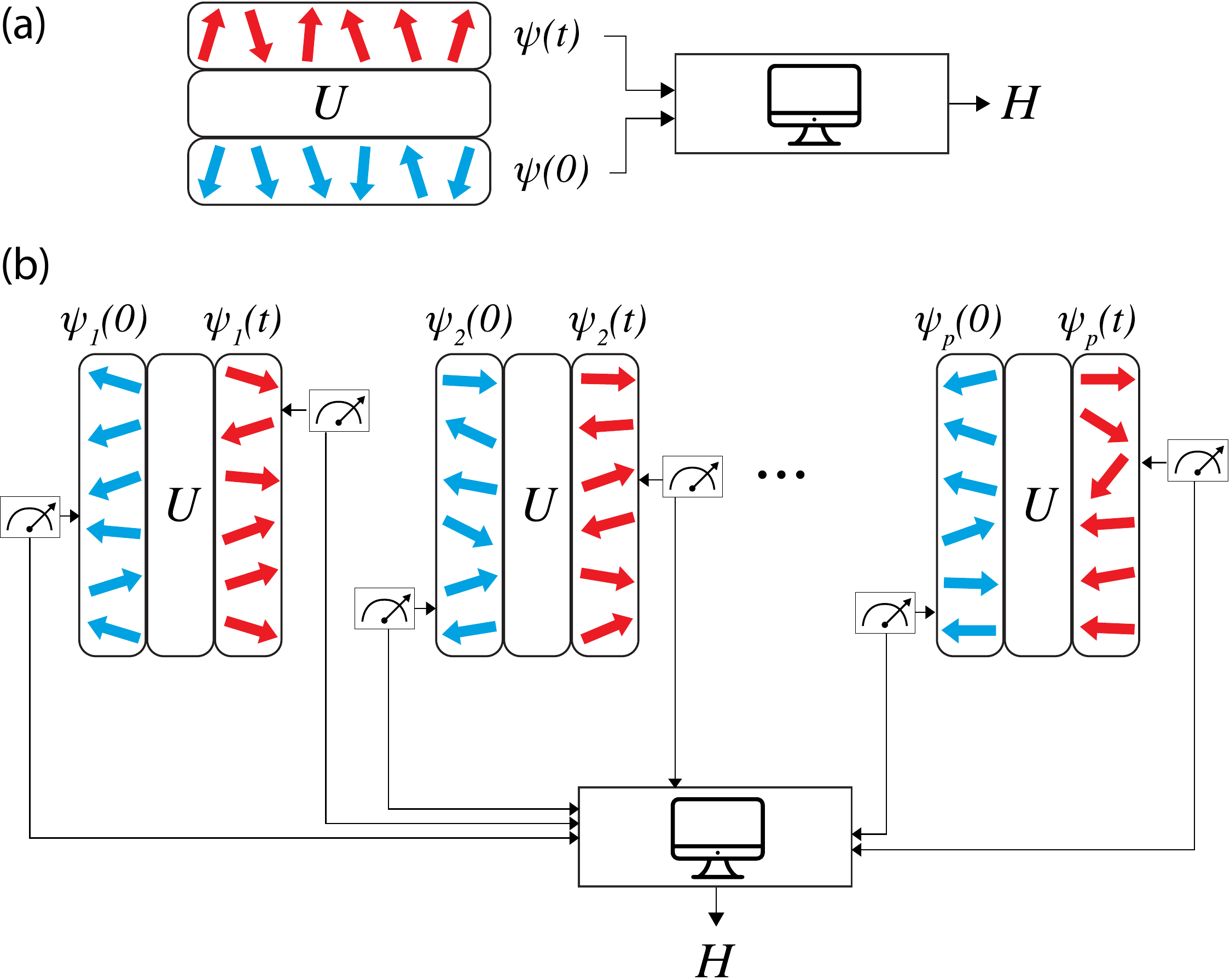}
    \caption{Two protocols for Hamiltonian tomography. (a) From a single pair of initial and final states related by time evolving with the Hamiltonian, $|\psi(t)\ra=U|\psi(0)\ra$ with $U=\exp\left(-iHt\right)$, a generic Hamiltonian can be uniquely reconstructed. (b) A practical version using $p$ pairs of initial and final states related by time evolving with the Hamiltonian and requiring only local measurements on the initial and final states.}
    \label{fig-summary}
\end{figure}

In this Letter, we introduce two protocols for Hamiltonian tomography (\reffig{fig-summary}). The first is more of conceptual interest: we show that it is possible to {\em uniquely} reconstruct a generic many-body Hamiltonian with local interactions given only a {\em single} pair of generic initial and final states connected by time evolving with the Hamiltonian. Our approach is partly motivated by the recent proposals of reconstructing a many-body Hamiltonian from a single eigenstate or steady state  \cite{Garrison2015, Qi2017, Chertkov2018, Bairey2018, flammia}. As in these studies, our protocol relies on the physical assumption that the Hamiltonian is local, which implies that the number of its parameters scales polynomially with system size. However, in contrast to these works, our approach does not require a steady state and instead relies precisely on how a generic state changes in time.

Despite being conceptually interesting, this first protocol is impractical and we thus propose a second, practical version that uses multiple pairs of initial and final states connected by time evolving with the Hamiltonian. This protocol only requires measuring a set of local observables in each pair of initial and final states. We bound the errors in the reconstructed Hamiltonian due to errors in both the measurements and the ansatz of the Hamiltonian, and the fidelity of the reconstruction can be increased to unity by using more pairs of initial-final states. 

We emphasize that our locality assumption does not necessarily have to be spatial locality; we only require that all interactions in the Hamiltonian involve a fixed number of degrees of freedom that is independent of system size.  Hence, our protocol applies to all analog and digital quantum simulators, from superconducting circuits to trapped ion platforms. We only require measurement of local observables, which can be done with high precision thanks to recent advances in single-site resolution.

{\it Tomography from a single quench.}---Given a Hamiltonian $H$ and an initial state $\ket{\psi(0)}$ of a $D$-dimensional Hilbert space, we ask whether one can determine $H$ from  only $\ket{\psi(0)}$ and $\ket{\psi(t)} = e^{-iHt}\ket{\psi(0)}$.  Without any restrictions, it is impossible to determine $H$ uniquely, since $H$ has $O(D^2)$ parameters while the given wave functions contain $O(D)$ parameters. However, a many-body Hamiltonian with local interactions has number of parameters scaling polynomially with system size, $L$, whereas $D$ is exponential in $L$, and thus reconstructing a local $H$ is possible in principle. 

Without loss of generality, we consider a traceless many-body Hamiltonian and decompose it into $n$ local interaction terms:
\begin{equation}\label{eq-localH}
    H=\sum_{\alpha=1}^nc_\alpha \mc{O}_\alpha,
\end{equation}
where $\{\mc{O}_\alpha\}$ are traceless local Hermitian operators, $\{c_\alpha\}$ are coupling constants, and $n$ is a polynomial of $L$ due to locality. Our goal is to determine $\{c_\alpha\}$ from $\ket{\psi(0)}$ and $\ket{\psi(t)} = e^{-iHt}\ket{\psi(0)}$.  

Our approach is based on the simple observation of generalized energy conservation:
\beq \label{eq-moment}
\bra{\psi(t)}H^m\ket{\psi(t)}=\bra{\psi(0)}H^m\ket{\psi(0)}
\eeq
for any positive integer $m$. These equations place many constraints on the variables $\{c_\alpha\}$. Indeed, substituting \refeq{eq-localH} (with $c$ replaced by $x$, a symbol for unknowns) into \refeq{eq-moment} yields (repeated indices summed over):
\begin{equation}\label{eq-polysystem}
    M^{(m)}_{\alpha_1\alpha_2\cdots \alpha_m}x_{\alpha_1}x_{\alpha_2}...x_{\alpha_m}=0,
\end{equation}
where 
\beqa\label{eq-Mdef}
    M^{(m)}_{\alpha_1\alpha_2\cdots \alpha_m}=&\bra{\psi(0)}\mc{O}_{\alpha_1}\mc{O}_{\alpha_2}\cdots \mc{O}_{\alpha_m}\ket{\psi(0)}\\
    &-\bra{\psi(t)}\mc{O}_{\alpha_1}\mc{O}_{\alpha_2}\cdots \mc{O}_{\alpha_m}\ket{\psi(t)}.
\eeqa
More explicitly, we have $M^{(1)}_{\alpha_1}x_{\alpha_1}=0, M^{(2)}_{\alpha_1\alpha_2}x_{\alpha_1}x_{\alpha_2}=0$, etc., which we refer to as first order, second order, etc. This constitutes a system of polynomial equations for $\{x_\alpha\}$ which can be used to determine $\{c_\alpha\}$ (and thus $H$).  Note that without knowledge of $t$, one can always rescale $H$ and $t$ to leave $Ht$ invariant, so $\{c_\alpha\}$ will be determined up to an overall multiplicative factor.  Hereafter we omit the ``up to a factor'' caveat.   

Although \refeq{eq-moment} is valid for any positive integer $m$, it can be shown  \cite{supp} that at most $(D-1)$ of these equations are independent. As long as $D-1\geqslant n$, 
our procedure uses the first $n$ equations ($m=1,2\cdots n$) to determine $\{c_\alpha\}$.  We find that if there exists one example of a Hamiltonian, $H_0$, that can be uniquely reconstructed from this procedure, then a generic Hamiltonian can be uniquely reconstructed from this procedure.
The proof of our claim is based on the analytic properties of resultants of polynomial equations, and we leave the technical details to the Supplementary Material  \cite{supp}.  There we also show that given the time $t$ as an additional input, a generic $H$ can be determined completely (including the overall scale).

Our procedure applies to generic Hamiltonians and initial states, and it can be immediately generalized for mixed initial states.  However, there will be fine-tuned cases for which it fails.  For example, if $|\psi(0)\ra$ is an eigenstate of $H$, \refeq{eq-moment} is trivially satisfied and we cannot obtain $H$ from it. Nevertheless, in this special case one can in principle employ the methods in Refs. 
 \cite{Garrison2015, Qi2017, Chertkov2018, Bairey2018, flammia} to determine the Hamiltonian. Furthermore, if $H$ admits a conserved quantity $Q$ that can be decomposed into the same set of local operators, $\{\mc{O}_\alpha\}$, then \refeq{eq-moment} yields any linear combination of $H$ and $Q$.

{\it Proof of concept.}---In order to demonstrate uniqueness for generic cases, one needs to show the existence of $H_0$ for arbitrary system size, which is a difficult problem. In this Letter, we present three physically relevant checks.

The first is an analytical check on a spin-$1/2$ chain. We consider a translationally invariant transverse-field Ising model with three random couplings:
\begin{equation}
        H=c_1\sum_i Z_iZ_{i+1}+c_2\sum_i X_i+c_3\sum_i Y_iY_{i+2},
\end{equation}
 in which $X,Y,Z$ are Pauli operators. We have added a next-nearest-neighbor interaction; otherwise, the problem is trivial and the first order (linear) equation in \refeq{eq-polysystem} is enough to determine $c_1/c_2$. Using a small $t$ approximation, it can be analytically shown  \cite{supp} that the first and second order equations have two solutions, only one of which satisfies the third order equation, and this unique reconstruction for small $t$ is sufficient to establish unique reconstruction for finite $t$.  We emphasize that the small $t$ approximation is not a requirement of our protocol; it is an intermediate step in establishing the uniqueness of reconstruction.

We also numerically check reconstruction using finite time evolution for  similar models. For example, we check the Ising model with transverse and longitudinal fields and the Heisenberg model for chains up to length $L=10$:
\begin{equation}
\begin{aligned}
        H&=c_1\sum_{i=1}^{L-1} Z_iZ_{i+1}+c_2\sum_{i=1}^L X_i+c_3\sum_{i=1}^L Z_i,\\
        H&=c_1\sum_{i=1}^{L-1} X_iX_{i+1}+c_2\sum_{i=1}^{L-1} Y_iY_{i+1}+ c_3\sum_{i=1}^{L-1} Z_iZ_{i+1}.
\end{aligned}
\end{equation}
We find similar uniqueness results as above. 

The third check is closer to a generic Hamiltonian. We consider a transverse field Ising model with random spatially varying couplings $c_{1,i}$ and $c_{2,i}$:
\begin{equation}
        H=\sum_{i=1}^L c_{1,i} X_i+\sum_{i=1}^{L-1} c_{2,i} Z_iZ_{i+1}
\end{equation}
Due to the high computational complexity [the number of coefficients in \refeq{eq-polysystem}
is exponential in $n$], we only check the case of $L=4$ (7 local operators in total) for illustration. Using the Gr\"{o}bner basis  \cite{Cox} technique, we find that $n=7$ polynomial equations indeed determine $\{c\}$ uniquely.

The above checks provide evidence for the existence of $H_0$ for various classes of Hamiltonians, which strongly suggests that a generic local many-body Hamiltonian can be uniquely reconstructed from a single pair of initial-final states related via time evolving with the Hamiltonian.

{\it Reconstruction from multiple quenches.}---Though the above protocol is conceptually interesting, it is impractical because the experimental and computational complexity are both exponential in $n$.  Hence, below we present a more practical method for Hamiltonian tomography, whose experimental and computational complexity is only a polynomial of $n$ and whose sensitivity to errors can be controlled.

Since the high complexity of the first approach arises from the higher order equations [$m\geqslant 2$ in \refeq{eq-polysystem}], we will only keep the linear equation $(m=1)$. In order to uniquely determine $\{c_{\alpha}\}$, we need at least $(n-1)$ linear equations.  Hence we use  $p \geqslant (n-1)$ pairs of initial and final states related by time evolving with $H$:
\begin{equation}
    \{\ket{\psi_i(0)}\to \ket{\psi_i(t)}=e^{-iHt}\ket{\psi_i(0)},~i=1,2\cdots p\},
\end{equation}
As a result, we have a linear system of equations $Mx=0$, 
where $M$ is a $p\times n$ matrix with entries
\begin{equation}\label{eq-defM} M_{i\alpha}=\dirac{\psi_i(0)}{\mc{O}_\alpha}{\psi_i(0)}-\dirac{\psi_i(t)}{\mc{O}_\alpha}{\psi_i(t)}.
\end{equation}

In principle it is sufficient to use $p=n-1$, and the kernel of $M$ will generically be one dimensional and equal to $\{c_\alpha\}$. (The uniqueness can be addressed by the same technique as for the single quench protocol.)
In contrast to the first approach, the number of measurements required is proportional to the number of coefficients in $M$, which is $O(n^2)$.

Inevitably there will be experimental errors in state preparation, time evolution, and measurements, and these will lead to errors in the reconstructed Hamiltonian.  We can mitigate these errors by utilizing more than $(n-1)$ pairs of initial-final states. Using $p> n-1$, we want to find the best fit of $\{x\}$ for $Mx=0$ when $M_{i\alpha}$ has random error.  We use the standard least squares method for a homogeneous system of equations: the best estimate of $\{x\}$ is given by the right singular vector of $M$ with the smallest singular value (or equivalently, the eigenvector of $M^TM$ with the smallest eigenvalue). A similar approach was used in Ref.  \cite{PhysRevX.9.041011} to find local integrals of motion; however, while that approach used different time slices from evolving a fixed initial state, our approach fixes one time and uses different pairs of initial and final states, which leads to more efficient use of experimental resources, as we will discuss below.

{\it Stability against measurement errors.}---To quantify the error in the reconstructed Hamiltonian $H=\sum_\alpha x_\alpha \mc{O}_\alpha$, we define
$\theta$ to be the angle between the vector of reconstructed coupling constants $x$ and the vector of actual coupling constants $c$, and correspondingly the fidelity $F=|\cos\theta|$ and reconstruction error $E=|\sin\theta|$. 

Consider an error model in which each matrix element $M_{i\alpha}$ has an additive error uniformly distributed between $\pm \epsilon$. Using both standard perturbation theory and nonperturbative results in statistical theory  \cite{cai2018}, we find that the average  reconstruction error is bounded:
\begin{equation}\label{eq-errorgap}
    E \leqslant C \sqrt{\frac{n}{p}}\frac{\epsilon}{\lambda}
\end{equation} 
for $\epsilon<\lambda$, where $C$ is a constant and $\lambda$ is the gap between the two lowest singular values of $\frac{1}{\sqrt{p}}M$ ($\lambda$ becomes independent of $p$ as $p\rightarrow \infty$)  \cite{supp}. 

\begin{figure}
    \centering
    \includegraphics[width=\linewidth]{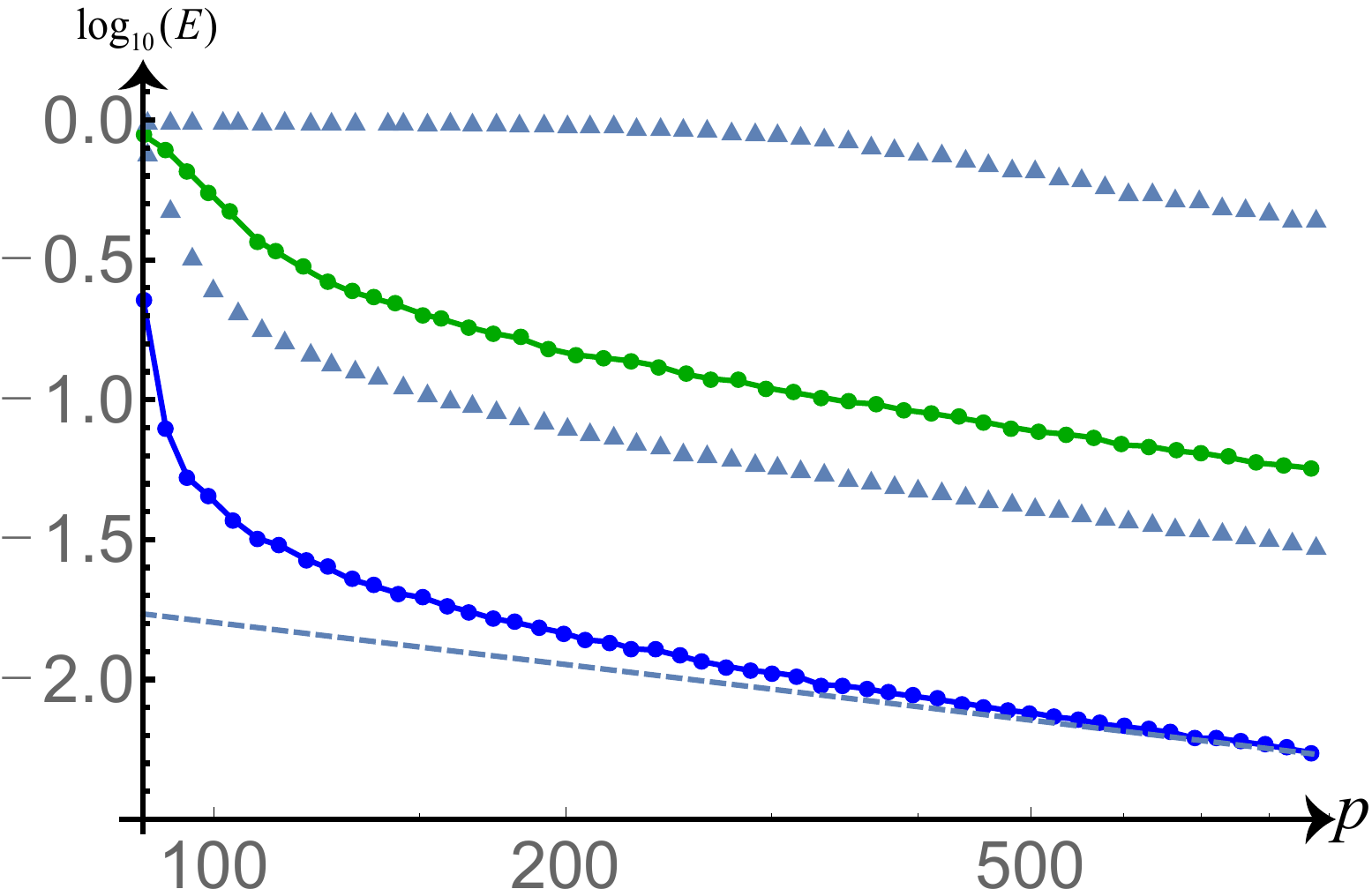}
    \caption{Reconstruction error vs number of initial and final pairs on log-log plot for $L=8,~t=1$ and error $\epsilon=0.1$ (green) and $\epsilon=0.01$ (blue).  The gray triangles denote reconstruction error from the procedure in Ref.  \cite{PhysRevX.9.041011}, using $p$ different time slices separated by $dt=1$ and one fixed initial state, for $\epsilon=0.1$ (top) and $\epsilon=0.01$ (middle).  Each point here is averaged over 200 realizations of error and random Hamiltonian \refeq{eq-Hrandomproduct} with couplings chosen from $(-1,1)$. The reference dashed line (bottom) corresponds to $E\propto\frac{1}{\sqrt{p}}$ predicted by \refeq{eq-errorgap}. Note that the initial decrease in error is steeper than the final rate.}
    \label{fig-fidelity_p}
\end{figure}
% This is caused by the dependence of $\lambda$ on $p$ when $p$ is small. 
Therefore, by increasing the number of pairs of initial-final states, $p$, one can decrease the reconstruction error \footnote{Why not fix $p$ and perform more measurements $k$ for each matrix element $M_{i\alpha}$ since the central limit theorem also guarantees that errors decay as $\frac{1}{\sqrt{k}}$? We find that in the experimentally relevant regime of $p$ marginally greater than $n$, error decreases faster than $\frac{1}{\sqrt{p}}$ (see \reffig{fig-fidelity_p}), and thus our approach is more efficient.}. As \refeq{eq-errorgap} implies, we would like $\lambda$ to be as large as possible. The gap is determined by $M$, which in turn depends on the important choices of time $t$ and the ensemble of initial states. Intuitively, larger $t$ is preferable so that the initial and final states are more distinguishable in terms of local observables. In the same vein, the initial states should not be too random; otherwise, the time evolution has little effect on changing local observables (for example, we find that Haar random initial states perform poorly).  At the same time, states in the initial ensemble should be distinct enough to provide independent information about $H$. 

We now discuss the dependence of $\lambda$ on these factors in more details.  To understand the singular values of $\frac{1}{\sqrt{p}}M$, we note that
\beqa\label{eq-covariance}
    \left(\frac{1}{p}M^TM\right)_{\alpha\beta}\!\!
    =\overline{\bra{\psi}\mc{O}_\alpha-\mc{O}_\alpha(t)\ket{\psi}\bra{\psi}\mc{O}_{\beta}-\mc{O}_{\beta}(t)\ket{\psi}},
\eeqa
where the overline denotes an average over the initial state ensemble.
We find that choosing the initial states from an ensemble of random product states provides a robust (and experimentally practical) scheme. Specifically, for a system of qubits we consider initial states given by
\begin{equation}
    \ket{\psi(0)}=\prod_i\ket{\phi_i},
\end{equation}
with $\ket{\phi_i}$ a random state on the Bloch sphere at site $i$.  (Alternatively, for each site one can choose $\ket{\phi_i}=|\pm 1\rangle$ in either the $X$,$Y$, or $Z$ basis.) Our results hereafter will assume the random Bloch sphere states as the initial states. 

\begin{figure}
    \centering
    \includegraphics[width=\linewidth]{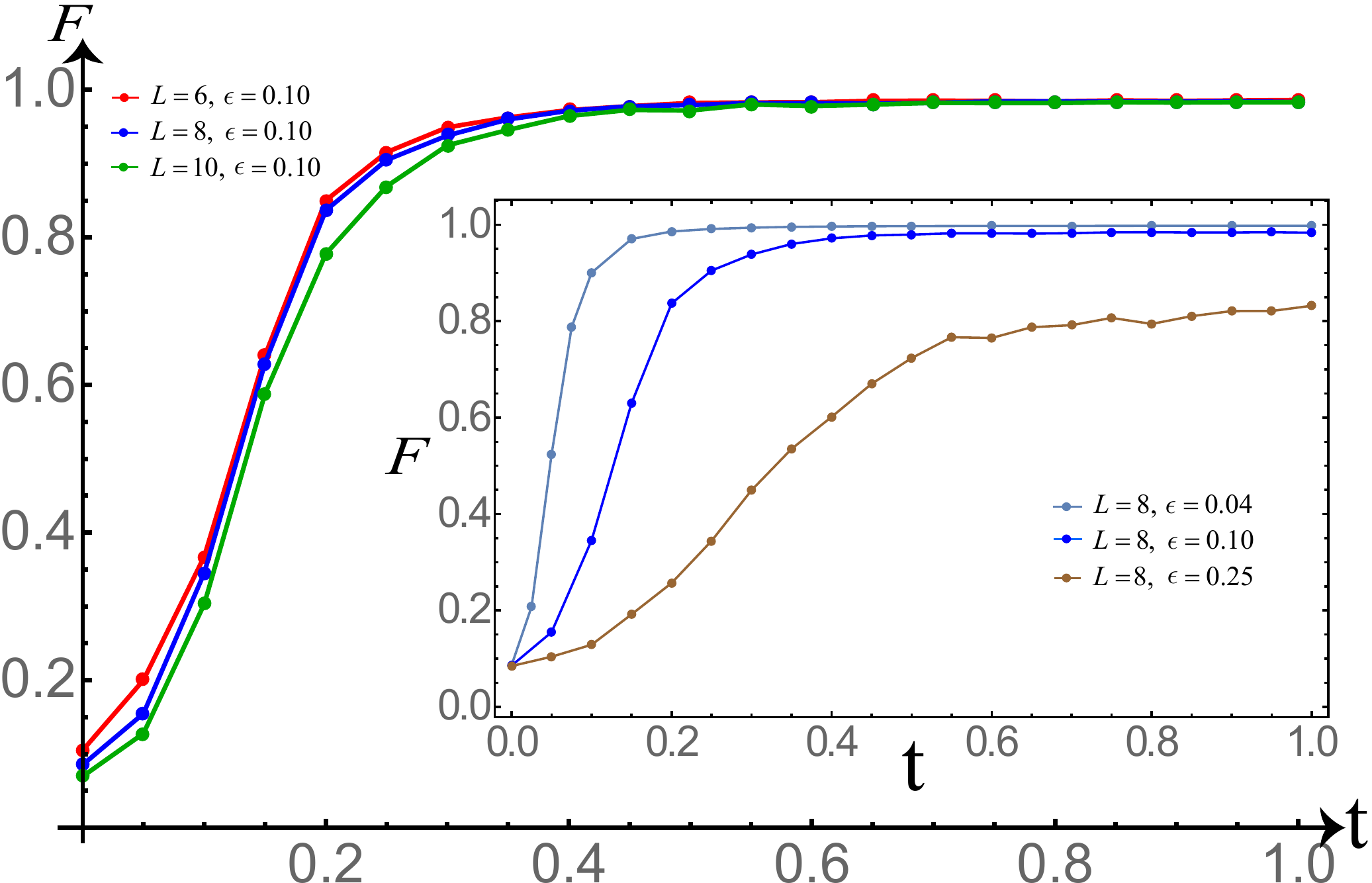}
    \caption{Reconstruction fidelity as a function of time interval $t$. We fix $p=2n$. Each point here is averaged over 200 realizations of error and random Hamiltonian \refeq{eq-Hrandomproduct} with couplings chosen from $(-1,1)$. In the main figure, three curves correspond to $L=6, 8, 10$ with errors in $(-0.1,0.1)$. In the subfigure, three curves correspond to errors in  $(-0.04,0.04)$, $(-0.1,0.1)$, and $(-0.25,0.25)$, with $L=8$.}
    \label{fig-fidelitytime}
\end{figure}
We analyze a generic class of Hamiltonians with random onsite and nearest-neighbor interactions in a spin$-1/2$ chain:
\begin{equation}\label{eq-Hrandomproduct}
    H= \sum_{i=1}^L\sum_{a=1}^3c_{ia}\sigma_i^a+\sum_{i=1}^{L-1}\sum_{a,b=1}^3 c_{iab}\sigma_i^a \sigma_{i+1}^b.
\end{equation}
Here $c_{ia}$ and $c_{iab}$ are random variables and $\sigma^{1,2,3} = \{X,Y,Z\}$ are Pauli operators. We will average over random Hamiltonian realizations $c_{ia},c_{iab} \in (-1,1)$ and error realizations in $(-\epsilon,\epsilon)$. In \reffig{fig-fidelity_p}, we plot the reconstruction error vs. $p$ and confirm the $1/{\sqrt{p}}$ dependence in \refeq{eq-errorgap}.  We use the experimentally reasonable timescale $t=1$ for $L=8$ and observe that $p=2n$ number of pairs is sufficient to achieve a high fidelity of 0.98 when $\epsilon=0.1$.

Figure \ref{fig-fidelity_p} also shows that our approach performs substantially better than evolving a fixed initial state with different times. This is because in the latter approach, if the times are too large, the final states will be locally thermal and hardly distinguishable.  On the other hand, states separated by small time are also hard to distinguish.  Using these states in \refeq{eq-defM} would result in $M$ and its gap $\lambda$ being smaller. This is also evident when we use our method and choose a small $t$ (see reconstruction fidelity versus $t$ in \reffig{fig-fidelitytime}; the fidelity at $t\approx 0$ is nonzero because the overlap between two random vectors in $\mathbb{R}^n$ is roughly $\frac{1}{\sqrt{n}}$). It is only at later times that $M$ and $\lambda$ become sufficiently nonzero, and the fidelity approaches unity. 

To understand the behavior of the fidelity curve in \reffig{fig-fidelitytime} and in particular its late time value, we have the following result.  Consider the correlation function $C_{\alpha\beta}(t)=\frac{1}{D}\Tr[\mc{O}_\alpha(0) \mc{O}_\beta(t)]$ (an $n\times n$ matrix) and let $l_{\max}$ be the maximum size of an operator in $\{\mc{O}_\alpha\}$.  For example, the operator $Z_i Z_j$ has size $2$ regardless of the distance $|i-j|$; a local Hamiltonian has $l_{\max}$ of $O(1)$.  Using \refeq{eq-covariance} and analyzing operator growth  \cite{supp}, we find 
\begin{equation}\label{gapbound}
    \lambda(t)\geqslant \left(\frac{1}{3}\right)^{l_{\max}/2}s_2\left(I-C(t)\right),
\end{equation}
where $s_2(I-C(t))$ denotes the 2nd smallest singular value of $I-C(t)$. Due to operator scrambling, we expect $C(t)$ to decay in time; assuming the eigenstate thermalization hypothesis (ETH),  matrix elements of $C(\infty)$ are $O(\frac{1}{L})$  \cite{Yongliang} with fluctuations exponentially small in system size  \cite{alhambra}.  We use similar techniques to find that $s_2(I-C(\infty))\geqslant 1-O(1/L)$.
\refeq{gapbound} thus provides an $O(1)$ lower bound for the gap. 

We emphasize that the above lower bound is independent of system size. Indeed, we find in Fig. \ref{fig-fidelitytime} that the late time reconstruction fidelity is insensitive to system size.  However, the time scale at which the maximal fidelity is reached depends on the time scale at which $C(t)$ asymptotes, which is expected to increase at most polynomially with system size. Note that the fidelity timescale also depends on the error magnitude: in the limit of zero error, the timescale for saturation approaches zero.  We observe in the numerics that for experimentally reasonable $L=10$, error $\epsilon=0.1$, and pairs of states $p=2n$, an $O(1)$ time is more than sufficient to reach maximal fidelity.

{\it Stability against errors in the ansatz of the Hamiltonian.}---In reality, interactions are not strictly compactly supported; there are small nonlocal interactions.  This motivates consideration of an enlarged Hamiltonian  
\begin{equation}\label{eq-realH}
    H=\sum_{\alpha} c_\alpha \mc{O}_\alpha+\sum_\beta c'_\beta \mc{O}'_\beta
\end{equation}
in which the support of $\mc{O}'_\beta$ is not necessarily bounded, but we assume $\norm{c'}\ll\norm{c}$.

Restricted by experimental and computational resources, it is desirable to model only the dominant interactions. Assuming the real Hamiltonian is described by \refeq{eq-realH} and we use our method to reconstruct an ``effective" Hamiltonian  $H=\sum_\alpha x_\alpha \mc{O}_\alpha$,
how different will $x$ be from $c$?

The matrix in \refeq{eq-defM} now consists of two pieces $(M;M')$ from $\{\mc{O}_\alpha\}, \{\mc{O}'_\beta\}$ respectively. The actual coefficient vector $\tvect{c}{c'}$ is the singular vector of $(M;M')$ with singular value 0. By restricting the operator set to $\{\mc{O}_\alpha\}$, one only measures $M$ and 
calculates its minimal singular vector $x$. 

We find  \cite{supp} that the reconstruction error between $x, c$ is controlled by
\begin{equation}
    E
    \leqslant
    \frac{\norm{\frac{1}{\sqrt{p}}M'}}{s_2\left(\sqrt{\frac{1}{p}}M\right)}
    \frac{\norm{c'}}{\norm{c}}.
\end{equation}
Recall that $s_2(\sqrt{\frac{1}{p}}M)$ denotes the second smallest singular value of $\frac{1}{\sqrt{p}}M$.
Using ETH, we show that $s_2(\sqrt{\frac{1}{p}}M) \geqslant(\frac{1}{3})^{l_{\max}/2}-O(\frac{1}{L})$, where $l_{\max}$ is the size of the largest operator in $\{\mc{O}_\alpha\}$ (not $\{\mc{O}'_\beta \}$).   Therefore $(\frac{1}{3})^{l_{\max}/2}$ is $O(1)$ and does not vanish even though the full Hamiltonian may contain arbitrarily nonlocal operators. We also show that  $\norm{\frac{1}{\sqrt{p}}M'}$ has an $O(1)$ bound. Thus, as long as $\norm{c'}\ll\norm{c}$, our reconstruction will succeed. 

{\it Summary.}---We have shown that a single quantum quench is sufficient in principle to reconstruct a generic many-body Hamiltonian with local interactions.  We also propose a practical version involving multiple quantum quenches from random initial product states and requiring only measurement of local observables. Using ETH, we analytically bound the reconstruction error arising from measurement errors and ignorance of nonlocal interactions.  The efficiency and robustness of our protocol enable quantum simulators to determine precisely the Hamiltonian being implemented.

{\it Acknowledgements.}---We thank Álvaro Alhambra, Ehud Altman, Xiaoliang Qi, and Beni Yoshida for helpful discussions, and Gabriela Secara for help with figure design. Z.L. is grateful for the hospitality of the Visiting Graduate Fellowship program at Perimeter Institute where this work was carried out. Z.L. is also supported by the PQI fellowship from Pittsburgh Quantum Institution. L.Z. is supported by the John Bardeen Postdoctoral Fellowship.  Research at Perimeter Institute is supported in part by the Government of Canada through the Department of Innovation, Science and Economic Development Canada and by the Province of Ontario through the Ministry of Colleges and Universities.

\bibliography{main.bib}

\end{document}

\title{Supplemental Material for ``Hamiltonian Tomography via Quantum Quench"}
\newcommand*{\PITT}{Department of Physics and Astronomy, University of Pittsburgh, Pittsburgh, Pennsylvania 15260, United States}
\newcommand*{\PQI}{Pittsburgh Quantum Institute, Pittsburgh, Pennsylvania 15260, United States}
\newcommand*{\PI}{Perimeter Institute for Theoretical Physics, Waterloo, Ontario N2L 2Y5, Canada}

\author{Zhi Li}\affiliation{\PITT}\affiliation{\PQI}\affiliation{\PI}
\author{Liujun Zou}\affiliation{\PI}
\author{Timothy H. Hsieh}\affiliation{\PI}

\begin{abstract}

\end{abstract}
	
\maketitle
	
\tableofcontents

\section{Single quench tomography}

\subsection{Dependence in the equal moments conditions}

To solve for $H$, we make use of the ``equal moments conditions", which we copy here for convenience:
\begin{equation}\label{eq-momentsupp}
   \bra{\psi(t)}H^m\ket{\psi(t)}=\bra{\psi(0)}H^m\ket{\psi(0)},~\forall m\in\mathbb{N}.
\end{equation}
Here, we show that at most $(D-1)$ of them are independent, where $D$ is the dimension of the Hilbert space.

We decompose $H$ as $H=\sum_i E_i\ket{i}\bra{i}$, then \refeq{eq-momentsupp} becomes:
\begin{equation}\label{eq-momentexpansion}
    \sum_{i=1}^D a_iE_i^m=0,
\end{equation}
where $a_i=|\braket{\psi(t)|i}|^2-|\braket{\psi(0)|i}|^2$. For simplicity, we first consider the case where $H$ has no degeneracy. In this case, the first $D$ equations ($m=0,1,\cdots,D-1$) are enough to show that $a_i=0~(\forall i)$. This is because the above equation can be viewed as a linear equation of $a_i$, whose coefficient matrix is the the Vandermonde matrix and its determinant is nonzero (provided that $H$ has no degeneracy):
\begin{equation}
\begin {vmatrix}
1 & E_1 & E_1^2 & \cdots & E_1^{D - 2} & E_1^{D - 1} \\
1 & E_2 & E_2^2 & \cdots & E_2^{D - 2} & E_2^{D - 1} \\
\vdots & \vdots & \vdots & \ddots & \vdots & \vdots \\
1 & E_D & E_D^2 & \cdots & E_D^{D - 2} & E_D^{D - 1}
\end {vmatrix}
=\prod_{1\leqslant i\leqslant j\leqslant D}(E_j-E_i)
\neq 0,
\end{equation}

Now that we know $a_i=0$, \refeq{eq-momentexpansion} holds for $\forall m\in\mathbb{N}$. Note that the equation with $m=0$ is redundant: $\sum_i a_i=0$ automatically due to the normalization condition $\braket{\psi(t)|\psi(t)}=\braket{\psi(0)|\psi(0)}=1$. Therefore, at most $(D-1)$ equations in \refeq{eq-momentsupp} are independent.

If $H$ has degeneracy, we can simply group $a_i$ together for those $i$'s with the same $E_i$. The same argument shows that at most $(\tilde{D}-1)$ of them are independent, where $\tilde{D}$ is the number of different eigenvalues of $H$ and $\tilde{D}\leqslant D$. 

In summary, at most $(D-1)$ equations in \refeq{eq-momentsupp} are independent. 

\subsection{Uniqueness of reconstruction}

Recall that we want to determine $\{c\}$ from \refeq{eq-momentsupp} with $m=1,2,\cdots,n$. More explicitly:
\begin{equation}\label{eq-supppolysystem}
 \begin{cases}
     &M^{(1)}_{\alpha_1}x_{\alpha_1}=0,\\
     &M^{(2)}_{\alpha_1\alpha_2}x_{\alpha_1}x_{\alpha_2}=0,\\
     &\cdots\cdots\\
     & M^{(n)}_{\alpha_1\alpha_2\cdots \alpha_n}x_{\alpha_1}x_{\alpha_2}...x_{\alpha_n}=0,
 \end{cases}
 \end{equation}
 where 
\beq\label{eq-Mdef}
    M^{(m)}_{\alpha _1\alpha _2\cdots \alpha _m}=\bra{\psi(0)}\mc{O}_{\alpha _1}\mc{O}_{\alpha _2}\cdots \mc{O}_{\alpha _m}\ket{\psi(0)}-
    \bra{\psi(t)}\mc{O}_{\alpha _1}\mc{O}_{\alpha _2}\cdots \mc{O}_{\alpha _m}\ket{\psi(t)}.
\eeq
Here we use $\{x\}$ as symbols of unknown variables, to distinguish it from $\{c\}$. 

In this section, we prove the following theorem:
 \begin{theorem}\label{thm-unique}
     Fix the system and fix a basis $\{\mc{O}_\alpha\}$ for local operators.  If there exists a Hamiltonian $H_0=\sum_\alpha c^0_\alpha \mc{O}_\alpha$ such that the coupling constants $c^0_\alpha$ can be uniquely determined on $\mathbb{C}$ (up to a multiplicative factor) from the above equations, then almost all $H=\sum_\alpha c_\alpha \mc{O}_\alpha$ can be uniquely determined (up to a factor).
 \end{theorem}
\begin{proof}
Without loss of generality, we assume $c_n^0\neq 0$. We will show that at most a  measure zero subset of $\mathbb{R}^n$ (the space of $c$) may fail to be determined uniquely. Since $\{c_n=0\}$ has measure zero, in the following we only need to consider the subspace $\mathbb{R}^n\backslash \{c_n=0\}$.

The polynomial system \refeq{eq-supppolysystem} contains $n$ equations for $n$ homogeneous variables, it is (formally) over-determined. However, by construction, it always has at least one nonzero solution $x=c$. Having at least one nonzero solution is equivalent to that a polynomial of the coefficients $M$, called Macaulay resultant, vanishes \cite{Gelfand}:
\begin{equation}
    \text{poly}_1(M)=0.
\end{equation} 

Here, we need a further result.
\begin{lemma}
When $c_n\neq 0$, then \refeq{eq-supppolysystem} has solutions other than $x=c$ (including those with $x_n=0$) if and only if some polynomials (denoted by poly$_2$ collectively) of the coefficients $\{M\}$ vanish simultaneously. 
\end{lemma}
\begin{proof}[Proof of Lemma]
Denote the original homogeneous equations as
\begin{equation}\label{eq-old}
   F^{(m)}(x_1,\cdots,x_n)=\sum_{\alpha_1\alpha_2\cdots \alpha_m}M_{\alpha_1\alpha_2\cdots \alpha_m}x_{\alpha_1}x_{\alpha_2}\cdots x_{\alpha_m} =0,~~(m=1,2,\cdots,n).
\end{equation} Under the following transformation:
\beqa\label{eq-subst}
    x_\alpha&\to y_\alpha+ c_\alpha y_n ~(\alpha=1,2,\cdots,n-1),\\
    x_n&\to  c_n y_n,\\
\eeqa
the original polynomial equations become
\begin{equation}
    G^{(m)}(y_1,\cdots,y_n)=F^{(m)}(y_1+ c_1y_n,\cdots,y_{n-1}+ c_{n-1}y_n, c_n y_n)=0,
\end{equation} 
which can be expanded as:
\begin{equation}\label{eq-new}
    G^{(m)}(y_1,\cdots,y_n)=\sum_{j=0}^m  G^{(m)}_j(y_1,\cdots,y_{n-1})y_n^{m-j},
\end{equation}
where $G^{(m)}_j$ is a homogeneous polynomial of $(y_1,\cdots,y_{n-1})$ with degree $j$. Solutions of \refeq{eq-old} are in one-to-one correspondence to solutions of \refeq{eq-new}, since the transformation \refeq{eq-subst} is invertible. The obvious solution $x=c$ of the original equations now becomes $y_1=\cdots=y_{n-1}=0$, $y_n\neq 0$, which imples:
\begin{equation}
    G^{(m)}_0(y_1,\cdots,y_{n-1})=0,\ \forall m.
\end{equation}
Therefore, we want to show that the following homogeneous equations
\begin{equation}\label{eq-newnew}
    G^{(m)}(y_1,\cdots,y_n)=\sum_{j=1}^m G^{(m)}_j(y_1,\cdots,y_{n-1})y_n^{m-j}
\end{equation}
has no solutions other than $[0,0,\cdots,0,1]$ if and only if the coefficients are special in the sense that some polynomial relations of the coefficients are satisfied.

Denote $\{a\}$ to be the set of coefficients (of $\{y\}$) in those $G^{(m)}(y)$. Denote $N$ to be the number of coefficients $\{a\}$. Now let us regard $G^{(m)}$ as polynomials of both $a$ and $y$ (linear in $a$, homogeneous in $y$), $G^{(m)}(a;y)$. Construct the ideal
\begin{equation}
    J=(G^{(1)}(a;y), G^{(2)}(a;y),....., G^{(n)}(a;y)),
\end{equation}
it defines a variety on $\mathbb{C}^N \times \mathbb{CP}^{n-1}$. 
Consider the ideal $I$ corresponding to $\mathbb{C}^N \times \{y_1=\cdots=y_{n-1}=0\}$, namely,
\begin{equation}
    I=(y_1,y_2,\cdots,y_{n-1}).
\end{equation}
We construct the ideal quotient  \cite{Cox} $(J:I^\infty)$. This construction serves to exclude $y_1=\cdots=y_{n}=0$ (which should be excluded since we are considering homogeneous variables) and the obvious solution $[0,\cdots,0,1]$. Then the variety $Z((J:I^\infty))$ is exactly the space of those extra solutions $(a;y)$, and the projection of $Z((J:I^\infty))$ to $\mathbb{C}^N$, denoted by $\pi(Z((J:I^\infty)))$, is exactly the space of ``special" coefficients.

According to the main theorem of elimination \cite{Cox}, $\pi(Z((J:I^\infty)))$ is Zariski closed (note that to use the main theorem of elimination, it is essential to work on $\mathbb{C}$ instead of $\mathbb{R}$, and to project along projective space $\mathbb{CP}^{n-1}$ instead of $\mathbb{C}^{n-1}$). In other words, it can be described as the set of common zeros for several polynomials of $a$.  Since each $a$ is a linear combination of original coefficients $M$ in \refeq{eq-old}, we conclude that \refeq{eq-old} has solutions other than $x\propto c$ if and only if some polynomial relations (``poly$_2$") of the coefficients $M$ are satisfied, at least when $ c_n\neq 0$.

\end{proof}

Back to the original problem. The success of uniquely reconstructing $H_0$ implies:
\begin{equation}\label{eq-nonvanish}
    \text{poly}_2(M(c^0, t, \psi(0)))\neq 0.
\end{equation}
According to \refeq{eq-Mdef}, each $M^{(m)}_{\alpha_1\alpha_2\cdots \alpha_m}$ is a real-analytic function of $\{c_\alpha\}, t$ and $|\psi(0)\ra$. Therefore poly$_2(M)$ is also real-analytic in $\{c_\alpha\}, t$ and $|\psi(0)\ra$. It can be shown \cite{Mityagin} that a real-analytic function is nonzero almost everywhere if it is nonzero at one point. Therefore, \refeq{eq-nonvanish} implies $\text{poly}_2(M(c, t, \psi(0)))\neq 0$ for generic $c$, which implies the success for uniquely reconstructing a generic $H$.
\end{proof}

\subsection{Determining the multiplicative factor}
% \begin{theorem}
% For generic $H$ and $\ket{\psi(0)}$, if $H$ is already known up to a factor, then \refeq{eq-problem} determines $H$ uniquely (provided $t$ is known).
% \end{theorem}
Suppose we know $H$ up to a multiplicative factor, we show that generically this factor can be determined from $\ket{\psi(t)}=e^{-iHt}\ket{\psi(0)}$ and the knowledge of $\ket{\psi(t)}$, $\ket{\psi(0)}$ and $t$. 

We will show that, equivalently, $e^{-iHt}\ket{\psi(0)}$ is generically injective in $t$ (even modulo the phase). Physically speaking, generically the Poincare recurrence cannot be exact. This means that, generically, given $|\psi(0)\ra$, $e^{-iHt}|\psi(0)\ra$ and $e^{-i\alpha  Ht}|\psi(0)\ra$ cannot be identical state, $\forall \alpha \neq 1$. So when $|\psi(t)\ra$ and $t$ are also given, $H$ can be generically determined without the freedom of multiplying a factor.

We expand $H$ as $H=\sum_i E_i\ket{i}\bra{i}$. Then:
\begin{equation}\label{eq-findscale}
    \ket{\psi(t)}=\sum_i e^{-iE_it} \braket{i|\psi(0)}\ket{i}.
\end{equation}
If $\ket{\psi(t)}\propto\ket{\psi(t')}$, then 
\begin{equation}
    e^{-iE_it}=e^{-iE_it'-i\theta}, \text{~for $i$ such that } \braket{i|\psi(0)}\neq 0,
\end{equation}
or equivalently,
\begin{equation}
    \left(E_i-\frac{\theta}{t-t'}\right)(t-t')=2\pi p_i,~p_i\in\mathbb{Z}, \text{~for $i$ such that } \braket{i|\psi(0)}\neq 0.
\end{equation}
This is impossible generically, since generically $\braket{i|\psi(0)}\neq 0$ for $\forall i$, and $E_i$ are mutually incommensurate even after shift.

The above analysis also provides us an algorithm to find the scale. Before presenting the algorithm, we note that any such algorithm will require precise information about $H$ and $\ket{\psi}$, and will be numerically unstable due to approximate Poincare recurrence. 

Assuming we know $H$ up to multiplicative factor, and $\psi(0), \psi(t)$ up to phases, then \refeq{eq-findscale} gives us:
\begin{equation}
    E_i t=i\log \frac{\braket{i|\psi(t)}}{\braket{i|\psi(0)}}\defeq \alpha_i+2\pi p_i+\theta,  \text{~for $i$ such that } \braket{i|\psi(0)}\neq 0.
\end{equation}
Denote $E_{ij}=E_i-E_j$ (and similarly for $\alpha_{ij}$ and $p_{ij}$). Picking up $i, j, k$ such that $\frac{E_{ij}}{E_{jk}}$ is irrational and \\ $\braket{i|\psi(0)}, \braket{j|\psi(0)}, \braket{k|\psi(0)}\neq 0$, then:
\begin{equation}
    p_{ij}E_{jk}-p_{jk}E_{ij}=\frac{E_{ij}\alpha_{jk}-E_{jk}\alpha_{ij}}{2\pi}.
\end{equation}
This is a $\mathbb{Z}-$linear combination of two  incommensurate numbers $E_{jk}$ and $E_{ij}$, therefore the coefficients can be found uniquely (by exhaustive  search) . Then $t$ can be determined from:
\begin{equation}
    E_{ij}t=\alpha_{ij}+2\pi p_{ij}.
\end{equation}

\subsection{Analytical checks of uniqueness}
\subsubsection{Linear approximation}
According to the Taylor expansion of $e^{-iHt}$, the coefficients in the above equations become:
\beq
    M^{(m)}_{\alpha _1\alpha _2\cdots \alpha _m}= it\bra{\psi(0)}[H,\mc{O}_{\alpha _1}\mc{O}_{\alpha _2}\cdots \mc{O}_{\alpha _m}]\ket{\psi(0)}+O(t^2)\defeq \bar{M}^{(m)}_{\alpha _1\alpha _2\cdots \alpha _m}+O(t^2).
\eeq
We want to take $\bar{M}$ as an approximation to $M$, and use this approximation to understand whether the original polynomial system, \refeq{eq-supppolysystem}, has a unique solution. 

First, note that \refeq{eq-supppolysystem} with $M$ substituted by $\bar{M}$ always has at least one solution, \ie $x=c$. This can be seen by fixing $x=c$ and then expanding \refeq{eq-supppolysystem} in $t$. For the uniqueness, the following statement justifies our linear approximation. It tells us that if the linearized equation has a unique solution, then the solution of the original equation is also unique.
\begin{claim}
If there exists a poly$_2$, denoted by $f$, such that $f(\bar{M}(t))\neq 0$ as a function of $t$, then $f(M(t))\neq 0$ for small enough $t$ and also for generic $t$.
\end{claim}
\begin{proof}
%Denote the coefficients for the degree-$m$ equation in \refeq{eq-old} collectively as $\{a_m\}$. We have the following simple observation: under rescaling $a_m\to\lambda_m a_m$,  number of solutions for \refeq{eq-old} remain unchanged. Therefore, if $f(\{a_1\},\cdots,\{a_m\},\cdots)=0$, then  $f(\{\lambda_1 a_1\},\cdots, \{\lambda_m a_m\},\cdots)=0$ for $\forall \lambda_1,\cdots,\lambda_m,\cdots$. 
We use the following simple observation: under rescaling $M\to\lambda M$, solutions of \refeq{eq-old} remain unchanged. Therefore, the set of poly$_2$ actually generates a homogeneous ideal and hence $f$ can be assumed to be homogeneous without loss of generality.

Due to the homogeneity of $f$, if we expand  $f(M(t))$ in $t$, the term with the lowest degree in $t$ will exactly be $f(\bar{M})$. Therefore, $f(\bar{M})\neq 0$ implies $f(M(t))\neq 0$ for small enough $t$, which also implies that $f(M(t))\neq 0$ for generic $t$ due to analyticity.
\end{proof}

\subsubsection{Translationally invariant systems}

Our formalism simplifies a lot for translationally invariant systems.  In this case, the ansatz for the Hamiltonian will be:
\beq \label{eq-trinvH}
H=\sum_{\alpha } c_\alpha \sum_{j=1}^L \mc{O}_{\alpha j},
\eeq
where $L$ is the number of sites, $\mc{O}_{\alpha  j}$ are local operators at site $j$. Our equal moments condition has the same form as \refeq{eq-supppolysystem} with $\mc{O}_\alpha $ now equal to $\sum_{j=1}^L \mc{O}_{\alpha  j}$ ($\alpha$ goes over local operator types):
\begin{equation}
    M^{(m)}_{\alpha _1\alpha _2\cdots \alpha _m}=\bra{\psi(0)}(\sum_{j=1}^L \mc{O}_{\alpha _1 j})(\sum_{j=1}^L \mc{O}_{\alpha _2 j})\cdots (\sum_{j=1}^L \mc{O}_{\alpha _m j})\ket{\psi(0)}-\bra{\psi(t)}(\sum_{j=1}^L \mc{O}_{\alpha _1 j})(\sum_{j=1}^L \mc{O}_{\alpha _2 j})\cdots (\sum_{j=1}^L \mc{O}_{\alpha _m j})\ket{\psi(t)}.
\end{equation}
The linear approximation will be determined by:
\begin{equation}
    \bar{M}^{(m)}_{\alpha _1\alpha _2\cdots \alpha _m}=it\bra{\psi(0)}\left[H,(\sum_{j=1}^L \mc{O}_{\alpha _1 j})(\sum_{j=1}^L \mc{O}_{\alpha _2 j})\cdots (\sum_{j=1}^L \mc{O}_{\alpha _m j})\right]\ket{\psi(0)}.
\end{equation}
\subsubsection{A spin chain example}
We consider the spin chain example in the main text:
\begin{equation}\label{eq-suppchain}
        H= c_1\sum_i Z_iZ_{i+1}+ c_2\sum_i X_i+ c_3\sum_i Y_i Y_{i+2}.
\end{equation}
In this model, there are three types of local operators, so we will try to use three equations to determine $c$. We will show that the linear approximation equations, with $\ket{\psi(0)}$ taken as a translationally invariant product state $\ket{\psi}=\prod_i\ket{n_i}$ ($\ket{n_i}$ is a fixed state on the Bloch sphere), indeed generically have a unique solution.

The linearized equations are as follows:
\beqa
    &\sum_{i=1}^L c_\delta\bra{\psi}\left[ \mc{O}_{\delta 1}, \mc{O}_{\alpha  i}\right]\ket{\psi}  x_\alpha =0,\\
    &\sum_{i=1}^L\sum_{j=1}^L c_\delta\bra{\psi}\left[\mc{O}_{\delta 1},\mc{O}_{\alpha i} \mc{O}_{\beta j}\right]\ket{\psi} x_\alpha  x_\beta =0,\\
    &\sum_{i=1}^L\sum_{j=1}^L\sum_{j=1}^L c_\delta\bra{\psi}\left[\mc{O}_{\delta 1},\mc{O}_{\alpha i} \mc{O}_{\beta j}\mc{O}_{\gamma k}\right] \ket{\psi}  x_\alpha  x_\beta x_\gamma=0.\\
\eeqa
In the above equations, $ c_\delta$ are constants, $x_\alpha, x_\beta, x_\gamma$  are variables, summation over $\alpha,\beta,\gamma,\delta$ are omitted. We only keep $\mc{O}_{\delta 1}$ instead of $\sum_l \mc{O}_{\delta l}$ due to translational invariance. Here, we will only briefly describe what happens instead of presenting the explicit form of the above equations\footnote{It is lengthy and not easy to work out by hand. We did it symbolically on computer.}.

For the 1st equation, note that only when local terms $\mc{O}_{\delta1}$ and $\mc{O}_{\alpha  i}$ have overlap can the commutator be nonzero, hence only a few terms contribute.

For the 2nd equation, commutators survive only if $\mc{O}_{\delta 1}$ has overlap with $\mc{O}_{\alpha  i}\mc{O}_{\beta j}$. There are two possibilities:
\begin{itemize}
    \item When one of $i$ and $j$ is far away from 1, say $j$, then it is just $[\mc{O}_{\delta 1},\mc{O}_{\alpha  i}]\mc{O}_{\beta j}$. This produces a term proportional to $L$ as well as an $O(1)$ ``offset'' (see below for an explicit example).  The term proportional to $L$ is exactly the one in the 1st equation.
    \item  When $\mc{O}_{\delta 1}$, $\mc{O}_{\alpha i}$ and $\mc{O}_{\beta j}$ are ``connected", we get $O(1)$ non-vanishing terms. 
\end{itemize} 
At the end, only $O(1)$ terms coming from the connected case and $O(1)$ terms coming from the offset of the disconnected case contribute.

For example, consider the case such that $\delta\leftrightarrow ZZ, \alpha\leftrightarrow ZZ, \beta\leftrightarrow X$, then 
\begin{equation}
    \left[Z_1Z_2, (\sum_i Z_iZ_{i+1})(\sum_i X_i)\right]=(\sum_i Z_iZ_{i+1})(iY_1Z_2+iZ_1Y_2),
\end{equation}
and
\beqa
    \bra{\psi}\left[Z_1Z_2, (\sum_i Z_iZ_{i+1})(\sum_i X_i)\right]\ket{\psi}
    &=\bra{\psi}\left(Z_0Z_1+Z_1Z_2+Z_2Z_3+\sum_{i\neq 0,1,2}Z_iZ_{i+1}\right)(iY_1Z_2+iZ_1Y_2)
    \ket{\psi}\\
    &= \underbrace{2(n_z^2n_x+n_x+in_yn_z)}_{\text{connected piece}}+(2L\underbrace{-6)n_z^2(in_yn_z)}_{\text{offset}}.
\eeqa
Here $n_x=\bra{n}X\ket{n}$, etc. The term proportional to $L$ is given by a corresponding term $\bra{\psi}[Z_1Z_2,(\sum_i X_i)]\ket{\psi}$ in the first equation.

For the 3rd equation, the situation is similar. The final equation contains $L^2, L^1$ and $L^0$ terms;  the $L^2$ term is exactly the 1st equation while the $L^1$ term can be derived from the 1st and 2nd equation. Only the $L^0$ term (from connected correlations and offsets) contribute.

We calculated the solution of the 1st and 2nd equation, and found that there are two solutions up to a factor. This is as expected, since a line  and a quadratic curve in two dimensions (two independent unknowns here) generically has two intersections. Moreover, it can be explicitly checked that only one of the solution survives the 3rd equation. 

In summary, we have analytically proved that our procedure works for the model in \refeq{eq-suppchain}.

\section{Multiple quench tomography}

\subsection{Homogeneous linear regression}
In this section, we consider the following homogeneous linear regression problem. Assume
\begin{equation}
    \sum_\alpha  M_{i\alpha }x_\alpha =0,
\end{equation}
for all $i$, but the knowledge of $M$ comes with errors. How should we get a good estimation for the vector $x$?

Let us apply the least square method to estimate $x$, \ie we will find $x$ which minimizes 
\begin{equation}\label{eq-penalty}
    S=\sum_i\text{error}_i^2=\sum_i(\sum_\alpha  M_{i\alpha }x_\alpha )^2=x^TM^TMx.
\end{equation}
However, since each equation is homogeneous, a constraint should be imposed to fix the overall magnitude of $x$ (otherwise the least square solution will be $x=0$). We use the following natural constraint:
\begin{equation}
    \sum_\alpha  x_\alpha ^2=x^2=1.
\end{equation}
This constraint can also be regarded as fixing $\Tr(H^2)$ if $\Tr(\mc{O}_\alpha  \mc{O}_{\alpha '})\propto\delta_{\alpha \alpha '}$.

Applying the method of Lagrange multipliers, we get the following equations:
\beqa
&M^TMx-\lambda x=0\\
&x^Tx=1,
\eeqa
where $\lambda$ is the Lagrange multiplier. Therefore, $x$ is an eigenvector of $M^TM$ with eigenvalue $\lambda$. Plug back into the original penalty function \refeq{eq-penalty}, we get:
\begin{equation}
    S=\lambda x^Tx=\lambda.
\end{equation}
Therefore, we should pick the vector $x$ corresponding to the minimal $\lambda$. In order words, the least square solution of the linear regression problem is the singular vector of $M$ corresponding to the minimal singular value $\sqrt{\lambda}$.

\subsection{Stability against error}
In this section we prove that the average reconstruction error caused by errors in $M$ is upper bounded by
\begin{equation}\label{eq-error}
    C\sqrt{\frac{n}{p}}\frac{\epsilon}{\lambda},
\end{equation} 
when $\epsilon$ is small enough. Here $\lambda$ is the gap between the smallest singular value (0) and the 2nd smallest singular value of $\frac{1}{\sqrt{p}}M$. In this subsection, we will assume that $M$ is gapped ($\lambda>0$), otherwise the above inequality holds trivially.

\subsubsection{Perturbative analysis}

We would like to compare the singular vectors of $M+E$ and $M$, where $E$ is a $p\times n$ matrix or errors. Only in this section, to simplify the computation, we model the error to be i.i.d $N(0,\epsilon)$. Due to properties of Gaussian distributions, elements of $U^\dagger EV$ are also iid $N(0,\epsilon)$ for any unitary matrices $U$  ($p$ by $p$) and $V$ ($n$ by $n$). Therefore, we can assume $M$ is in its singular-value-decomposed form:
\begin{equation}\label{eq-MSVD}
\left[\begin{array}{ccccc}
\sqrt{p}\lambda_1&         &      &       \\
& \sqrt{p}\lambda_2    &  \\
&   &     \sqrt{p}\lambda_{n-1}  &\\
&   &     &0  \\
\hline
0& 0  & 0    &0\\
\end{array}\right].
\end{equation}

Now we perform standard perturbation theory on $M^TM$. Since
\begin{equation}
    \Delta(M^TM)=(M+E)^T(M+E)-M^TM=M^TE+E^TM+E^TE,
\end{equation}
Let $\ket{1}$ be the eigenvector of $M^TM$ with eigenvalue $0$.  Then the perturbed eigenvector is:
\begin{equation}
    \ket{\bar{1}}=\ket{1}-\sum_{i=2}^{n}\frac{\bra{i}M^TE+E^TM\ket{1}}{p\lambda_i^2}\ket{i}+O(\epsilon^2).
\end{equation}
Using \refeq{eq-MSVD} and $M\ket{1}=0$, we have:
\begin{equation}
    \bra{i}M^TE+E^TM\ket{1}=\bra{i}M^TE\ket{1}=\sqrt{p}\lambda_i\epsilon_i,
\end{equation}
where $\epsilon_i=\bra{i}E\ket{1}$ is a $N(0,\epsilon)$ random variable. Therefore,
\begin{equation}
    \ket{\bar{1}}=\ket{1}-\sum_{i=2}^{n}\frac{\epsilon_i}{\sqrt{p}\lambda_i}\ket{i}+O(\epsilon^2),
\end{equation}
and the angle between reconstructed and actual couplings is given by 
\beqa
    |\sin\theta|&=\norm{\sum_{i=2}^{n}\frac{\epsilon_i}{\sqrt{p}\lambda_i}\ket{i}}+O(\epsilon^2)\\
    &=\sqrt{\sum_{i=2}^{n}\frac{\epsilon_i^2}{p\lambda_i^2}}+O(\epsilon^2).
\eeqa
On average, we will have 
\begin{equation}
    \overline{|\sin\theta|}\leqslant\sqrt{\frac{n}{p}}\frac{\epsilon}{\bar{\lambda}}+O(\epsilon^2),
\end{equation}
where $\bar{\lambda}^2$ is the harmonic mean of $\lambda_i^2$, which is no less than the minimal one ($\lambda$). Therefore the expectation value of error $\sin\theta$ is bounded by \refeq{eq-error}.

\subsubsection{Nonperturbative analysis}

Strictly speaking, one needs to worry about the higher order terms in the perturbation theory, even if the perturbation theory converges. It is also possible that when $\epsilon$ is some $o(1)$ value (as $n\to\infty$ or $p\to\infty$), the 1st order result is already not a good approximation. Therefore, we need some nonperturbative results to estimate the error in the realistic case, where $\epsilon$ is at constant level (albeit small).

The problem of singular vector perturbation has been actively studied (nonperturbatively) in the mathematical and statistical literature. Here we point out the following result which is relevant to us, see theorem 3 in  Ref.~\onlinecite{cai2018}:
\begin{equation}
    \mathbb{E}\norm{\sin\Theta(V,V')}^2\leqslant \frac{Cn(p\lambda^2\epsilon^2+p\epsilon^4)}{p^2\lambda^4}\leqslant\frac{2Cn}{p}\frac{\epsilon^2}{\lambda^2} ~(\text{if~} \epsilon\leqslant\lambda).
\end{equation}
Here,  $C$ is some absolute constant; $V$ and $V'$ are the space spanned by the first $n-1$ (right) singular vectors of $M$ and $M+E$, respectively, therefore $\norm{\sin\Theta(V,V')}=|\sin\theta|$.

\subsection{Analysis of the gap}

Recall that
\begin{equation}
    M_{i \alpha  }=\bra{\psi_i}\mc{O}_\alpha  -\mc{O}_\alpha  (t)\ket{\psi_i},
\end{equation}
where $i=1\cdots p$ label the initial-final states pair, $\alpha=1\cdots n$ label the parameters in $H$. Define a random variable
\begin{equation}
    X=(\bra{\psi}\mc{O}_1-\mc{O}_1(t)\ket{\psi}, \cdots, \bra{\psi}\mc{O}_n-\mc{O}_n(t)\ket{\psi}).
\end{equation}
It is a random variable since $\ket{\psi}$ comes from a random ensemble. Then the matrix elements of $\frac{1}{p}M^TM$ are
\begin{equation}\label{eq-kMM}
\frac{1}{p}(M^TM)_{\alpha \beta}=\frac{1}{p}\sum_{i=1}^p\bra{\psi_i}\mc{O}_{\alpha}-\mc{O}_{\alpha}(t)\ket{\psi_i}\bra{\psi_i}\mc{O}_\beta -\mc{O}_\beta (t)\ket{\psi_i}=\frac{1}{p}\sum_{i=1}^p X_i^T X_i.
\end{equation}
Therefore, $\frac{1}{p}M^TM$ is the sample covariance matrix (not centered) of the random variable $X$,  or in other words the sample average of $X^T X$.

In the following, we will use the actual covariance matrix of $X$ as an estimation of $\frac{1}{p}M^TM$:
\begin{equation}\label{eq-approx}
    \frac{1}{p}M^TM \leftarrow \mathbb{E}[X^TX]=\overline{\expct{\mc{O}_\alpha -\mc{O}_\alpha (t)}\expct{\mc{O}_\beta -\mc{O}_\beta (t)}},
\end{equation}
where the overline means ensemble average over initial states $\ket{\psi}$. In other words, we take $p\to\infty$ limit. The behavior at finite $p$ will be discussed in subsection \ref{sec-finitetp}.

We are free to choose the initial states ensemble. But in any case, when $t=0$ we have $M=0$ and the gap of $M/\sqrt{p}$ is $\lambda=0$. This is why at short times the fidelity is low.

\subsubsection{Haar random initial states}
In this case, a crude estimation will show that the stability is bad. Indeed, both the harmonic mean gap $\bar{\lambda}$ and the minimal gap are bounded by the arithmetic mean of the eigenvalues:
\begin{equation}
    \lambda^2\leqslant \bar{\lambda}^2\leqslant\frac{1}{n-1}\Tr{(\frac{1}{p}M^TM)}=\frac{1}{n-1}\sum_{\alpha=1}^n \overline{\bra{\psi}\mc{O}_\alpha-\mc{O}_\alpha  (t)\ket{\psi}^2},
\end{equation}
With Haar integral,
\beqa\label{eq-HaarL}
    \overline{\bra{\psi}\mc{O}_\alpha -\mc{O}_\alpha  (t)\ket{\psi}^2}
    &= \int d\psi\bra{\psi}\mc{O}_\alpha  -\mc{O}_\alpha  (t)\ket{\psi}^2\\
    &=\frac{1}{D(D+1)}\left[\Tr^2(\mc{O}_\alpha  -\mc{O}_\alpha  (t))+\Tr[(\mc{O}_\alpha  -\mc{O}_\alpha  (t))^2]\right]\leqslant\frac{4}{D+1}.
\eeqa
Here we have used a Haar average formula and the fact that $\mc{O}_\alpha$ is traceless. This exponentially small gap makes the reconstruction very sensitive to the error.

\subsubsection{Constraint on operator growth from energy conservation}

Assuming $H$ has generic spectrum, then the late time value of the coefficient of $A$ in the expansion of $B(t)$ is:
\begin{equation}
\begin{aligned}
    \lim_{T\to\infty}\frac{1}{T}\int_0^T \frac{1}{D}\Tr(AB(t))dt&= \lim_{T\to\infty}\frac{1}{T}\int_0^T\frac{1}{D}\sum_{E,E'}
    \dirac{E}{A}{E'}e^{iE't}\dirac{E'}{B}{E}e^{-iEt}dt\\
    &=\frac{1}{D}\sum_E A_E B_E.
\end{aligned}
\end{equation}
with $A_E\equiv\la E|A|E\ra$ and $B_E\equiv\la E|B|E\ra$. For traceless local operators $A,B$, we will derive a formula for this quantity for systems with geometrically local interaction, assuming eigenstate thermalization hypothesis (ETH), following Ref.~\cite{Yongliang}.

ETH tells us that there exist smooth enough functions $f_A$ and $f_B$ such that:
\begin{equation}
    |A_E-f_A(\frac{E}{L})|\leqslant\frac{1}{\text{poly}(L)},
\end{equation}
and similarly for $B$, where $L$ is the number of sites (volume), and poly$(L)$ means some polynomial of $L$ with high enough degree.
Moreover, it can be shown that for a trace-less local operator $A$ \cite{Yongliang}:
\begin{equation}
    f_A(0)=0,~~\frac{f'_A(0)}{L}=\frac{\Tr(HA)}{\Tr(H^2)}, 
\end{equation}
and similarly for $B$. Therefore,
\begin{equation}\label{eq-overlapwithH}
\begin{aligned}
    &\frac{1}{D}\sum_E A_E B_E\\
    =&\frac{1}{D}\sum_{E} f'_A(0)\frac{E}{L} f'_B(0)\frac{E}{L}+\frac{1}{D}\sum_{E\leq L^{\frac{1}{2}+\epsilon}} \left(A_EB_E-f'_A(0)\frac{E}{L} f'_B(0)\frac{E}{L}\right)+\frac{1}{D}\sum_{E>L^{\frac{1}{2}+\epsilon}}\left(A_E B_E-f'_A(0)\frac{E}{L} f'_B(0)\frac{E}{L}\right)\\
    =&\frac{1}{D}\frac{\Tr(HA)}{\Tr(H^2)}\frac{\Tr(HA)}{\Tr(H^2)}\sum_E E^2+O(\frac{1}{L^2})\\
    =&\frac{\Tr(HA)\Tr(HB)}{D\Tr(H^2)}+O(\frac{1}{L^2}).
\end{aligned}
\end{equation}
The fact that the error is $O(\frac{1}{L^2})$ can be proved with some technical results in Ref.~\cite{Yongliang}. The first term here is $O(\frac{1}{L})$. For example, if $A=\mc{O}_\alpha, B=\mc{O}_\beta $, $H=\sum_{\alpha=1}^n   c_\alpha  \mc{O}_\alpha$, and $\Tr(\mc{O}_\alpha \mc{O}_\beta )\propto \delta_{\alpha \beta }$, then the first term is just:
\begin{equation}\label{eq-finalvalue}
\frac{ c_\alpha  c_\beta }{\sum_{\alpha=1}^n  c_\alpha ^2}.
\end{equation}

\refeq{eq-overlapwithH} tells us that if both $A$ and $B$ have overlap with the Hamiltonian, then the weight of $A$ in $B(t)$ will be only polynomially small, $O(\frac{1}{L})$, instead of exponentially small if one were to naively replace the time evolution operator by Haar random unitary. This can be intuitively understood from energy conservation: an $O(1)$ perturbation of $A$ gives an $O(1)$ perturbation to the conserved energy. This perturbation is then evenly distributed among each $B(t)$, so that the correlation of $A$ and $B(t)$ is of order $O(\frac{1}{L})$.

Regarding the deviation between $C(t)$ and the ETH value in \refeq{eq-overlapwithH}, denoted by $\frac{1}{D}\Tr(AB(\infty))$, it is proved in Ref.~\cite{alhambra} that:
\begin{equation}\label{eq-fluc}
     \lim_{T\to\infty}\frac{1}{T}\int_0^T \Big(\frac{1}{D}\Tr(AB(t))-\frac{1}{D}\Tr(AB(\infty))\Big)^2dt\leqslant\frac{1}{D}\norm{A}^2\norm{B}^2.
\end{equation}
We note that the only assumption used to prove this inequality is just a generic energy spectrum, rather than the more stringent ETH \cite{alhambra}.

\subsubsection{Random Product States: Exact Results}

Suppose at $t=0$ all $\mc{O}_\alpha $'s are Pauli strings normalized such that $\Tr(\mc{O}_\alpha (0)\mc{O}_\beta (0))=D\delta_{\alpha \beta }$. Let us expand $\mc{O}_\alpha (t)$ as
\begin{equation}\label{eq-operatorexpansion}
    \mc{O}_\alpha (t)=\sum_s P_s C_{s,\alpha}(t),
\end{equation}
where $P_s$ means Pauli operator with string $s$, \eg $s=(\cdots,I,I,Z,X,I,I,\cdots)$. Note $s\neq (\cdots,I,I,I,\cdots)$ because $\mc{O}_\alpha $ is taken to be traceless. The coefficients $ C_{s,\alpha}(t)$ depends on $s,i,t$, and $c$. Due to energy conservation, $H(t)=H$. So for each $s$, we have:
\begin{equation}\label{eq-conservationC}
    \sum_{\alpha}   C_{s,\alpha}(t) c_\alpha =\sum_{\alpha}   C_{s,\alpha}(0) c_\alpha =
        \begin{cases}
     c_\alpha, & s=\mc{O}_\alpha  \text{~for some~}\alpha\\
    0,& s\notin\{\mc{O}\}.
\end{cases}.
\end{equation}
Also note that $\Tr(\mc{O}_\alpha (0)\mc{O}_\beta (0))=D\delta_{\alpha \beta }$ implies
\begin{equation}\label{eq-normalization}
    \sum_s  C_{s,\alpha}(t) C_{s,\beta}(t)=\delta_{\alpha \beta }.
\end{equation}

Now consider the covariance:
\begin{equation}
\frac{1}{p}(M^TM)_{\alpha \beta }=\overline{\expct{\mc{O}_\alpha -\mc{O}_\alpha (t)}\expct{\mc{O}_\beta -\mc{O}_\beta (t)}}.
\end{equation}
We note the following formula for random product states average:
\begin{equation}
    \overline{\expct{P_s}\expct{P_{s'}}}=\delta_{ss'}
        (\frac{1}{3})^{l_s},~~~(l_s=\text{size of~}P_s).
\end{equation}
Here $P_s$ means a Pauli string. Alternatively, one can also use a random $XYZ$ ensemble. Namely, each spin can choose from $X=\pm 1, Z=\pm 1, Z=\pm 1$ with equal ($\frac{1}{6}$) probability. The above formula still holds for this ensemble.
With this formula, the covariance can be expanded as:
\begin{equation}\label{eq-MMexpansion}
\begin{aligned}
    \frac{1}{p}(M^TM)_{\alpha \beta }&=(\frac{1}{3})^{l_\alpha }\delta_{\alpha \beta }-(\frac{1}{3})^{l_\beta } C_{\beta,\alpha}(t)-(\frac{1}{3})^{l_\alpha } C_{\alpha,\beta}(t)+\sum_s  C_{s,\alpha}(t) C_{s,\beta}(t)(\frac{1}{3})^{l_s}\\
    &=\Big[(\frac{1}{3})^{l_\alpha }\delta_{\alpha \beta }-(\frac{1}{3})^{l_\beta } C_{\beta,\alpha}(t)-(\frac{1}{3})^{l_\alpha } C_{\alpha,\beta}(t)+\sum_{s\in\{\mc{O}\}}  C_{s,\alpha}(t) C_{s,\beta}(t)(\frac{1}{3})^{l_s}\Big]+\sum_{s\notin\{\mc{O}\}}  C_{s,\alpha}(t) C_{s,\beta}(t)(\frac{1}{3})^{l_s}\\
    &\defeq A_{\alpha \beta }+\tilde{A}_{\alpha \beta }.
\end{aligned}
\end{equation}
Here $\{\mc{O}\}$ is the set of local operators in the Hamiltonian.

Define three matrices $B,C(t),\tilde{C}(t)$ as: 
\beqa
    &B_{\alpha \beta }=(\frac{1}{3})^{l_\alpha }\delta_{\alpha \beta },\\
    & C_{s \alpha }(t)=\text{expansion coefficients in \refeq{eq-operatorexpansion}~~} (s\in \{\mc{O}\})\\
    & \tilde{C}_{s\alpha}(t)=\text{expansion coefficients in \refeq{eq-operatorexpansion}~~} (s\notin \{\mc{O}\}).
\eeqa
Then 
\beqa\label{eq-ACB}
    A(t)&=B-C(t)^TB-BC(t)+C(t)^TBC(t)\\
    &=(I-C(t)^T)B(I-C(t))
\eeqa
Therefore, $A(t)$ is a positive semi-definite matrix. Similarly, $\tilde{A}(t)=\tilde{C}(t)^TB\tilde{C}(t)$ is also positive semi-definite. 

Moreover, $A$ has the same zero mode as $\frac{1}{p}M^TM$. Indeed, since $\frac{1}{p}M^TM=A+\tilde{A}$ with both term positive semi-definite, a zero mode of $\frac{1}{p}M^TM$ must be a zero mode of $A$. Conversely, if $Ax=0$, then we must have $x=Cx$ since $B$ is positive definite. This also implies $\tilde{C}x=0$ since \refeq{eq-normalization} is just
\begin{equation}\label{eq-CCI}
    C^TC+\tilde{C}^T\tilde{C}=I.
\end{equation}

We will use the following result\footnote{This result is an easy corollary of Weyl's inequality, or can be proved by the Courant min-max principle.}: if $X\geqslant Y$ as Hermitian operators (in the sense that $X-Y$ is positive semidefinite), then
\begin{equation}
    \sigma_i(X)\geqslant \sigma_i(Y)
\end{equation}
for all $i$, where $\sigma_i(X)$ and $\sigma_i(Y)$ are the $i$-th smallest eigenvalues of $X$ and $Y$, respectively.
Since $\tilde{A}$ is positive semi-definite, we have: 
\begin{equation}
    A+\tilde{A}\geqslant A\geqslant  (\frac{1}{3})^{l_{\max}}(I-C^T)(I-C),
\end{equation}
where $l_{\max}$ is maximal length of Pauli operators that appear in $\{\mc{O}_\alpha \}$. Therefore
\begin{equation}
    \lambda^2=\sigma_{2}(A+\tilde{A})\geqslant\sigma_{2}\left((\frac{1}{3})^{l_{\max}}(I-C^T)(I-C)\right).
\end{equation}

Note that we have actually proved the existence of a gap of $M$, therefore the uniqueness of the reconstruction, without referring to ``generic". This is a consequence of ETH. Indeed, if there are other conserved quantities (and our approach for Hamiltonian tomography will not yield a unique solution), the ETH does not hold anymore.

\subsubsection{Random Product States: Understanding from ETH}
If we use the late time value \refeq{eq-finalvalue} predicted by ETH, and ignore the $O(\frac{1}{L^2})$ term, then:
\begin{equation}
    C	\leftarrow \frac{cc^T}{\norm{c}^2}\defeq  C_0.
\end{equation}
The spectrum of $(I-C^T_0)(I- C_0)$ will just be $0,1,\cdots,1$, where the $n-1$ eigenvectors with eigenvalue $1$ are orthogonal to $c$, and $c$ is the last eigenvector with eigenvalue $0$. Then we have: 
\begin{equation}
    \lambda^2\geqslant (\frac{1}{3})^{l_{\max}}.
\end{equation}
Taking the $O(\frac{1}{L^2})$ correction and the exponentially small fluctuation \refeq{eq-fluc} into consideration, we obtain:
\begin{equation}\label{eq-gaplowerbound}
    \lambda\geqslant (\frac{1}{3})^{l_{\max}/2}-O(\frac{1}{L}).
\end{equation}

As a side note, we may want to take $(I-C^T_0)B(I- C_0)$ as an approximation to $\frac{1}{p}M^TM$. Recall that $\frac{1}{p}M^TM=A(t)+\tilde{A}(t)$, this approximation amounts to ignore $\tilde{A}(t)$, $A(t)-A(\infty)$, and the $O(\frac{1}{L^2})$ correction in the end value. It turns out the approximation works quite well. 

\begin{figure}
    \centering
    \includegraphics[width=0.5\linewidth]{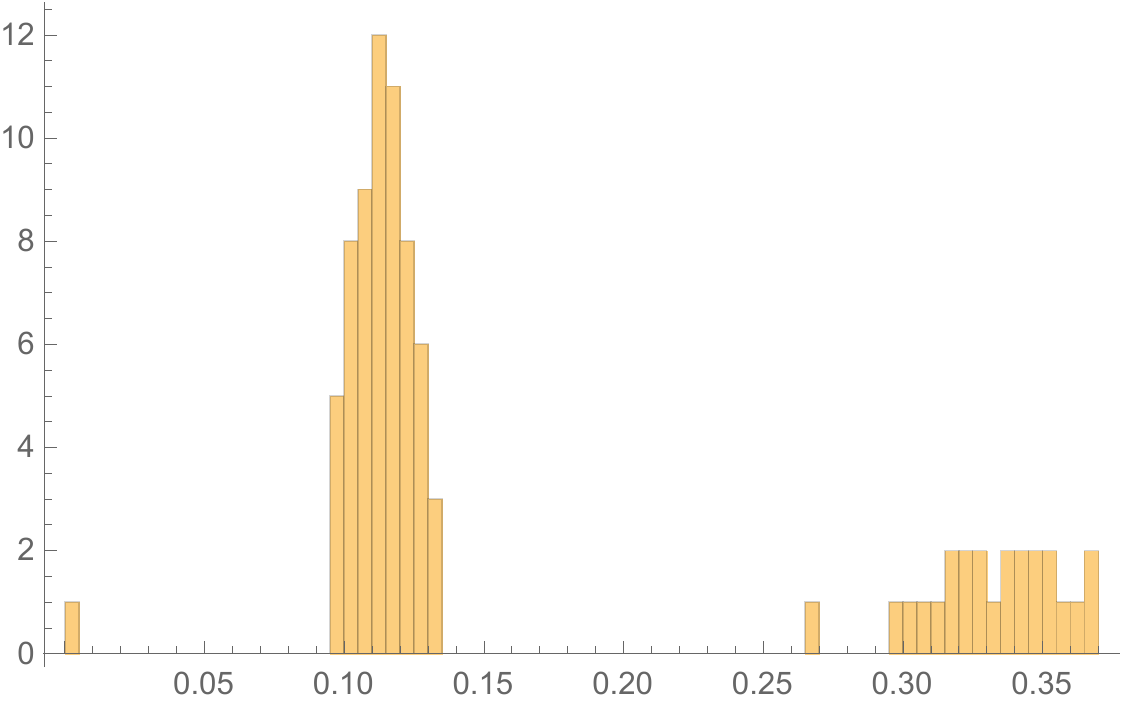}
    \caption{Histogram of eigenvalues of $\frac{1}{p}M^TM$. Here we only calculate the spectrum for one realization of $H$. $L=8$, $t=10$, $p=15,000$. }
    \label{fig-concentration}
\end{figure}

In Fig.~\ref{fig-concentration}, we plot the histogram of eigenvalues of $\frac{1}{p}M^TM$. In this example, we use random 1-body and 2-body Pauli operators as the basis $\{\mc{O}_\alpha \}$, so $l=1,2$. The matrix $(I-C^T_0)B(I- C_0)$ can be diagonalized exactly, and the spectrum is:
\begin{equation}
    0,~\frac{1}{9},\cdots,\frac{1}{9},~\frac{1}{3}-\frac{2}{9}{\mu}_1^2,~\frac{1}{3},\cdots,\frac{1}{3}.
\end{equation}
Here ${\mu}_1$ is the square sum of those $ c_\alpha $ corresponding to operators $\mc{O}_\alpha $ with length 1. Note that $\frac{1}{3}-\frac{2}{9}{\mu}_1^2\in[\frac{1}{9},\frac{1}{3}]$, the gap $\lambda^2$ is always $\frac{1}{9}$, as expected. The above spectrum concentrates at $\frac{1}{9}$ and $\frac{1}{3}$, with one exception at 0 the other one between  $(\frac{1}{9},\frac{1}{3})$. As can be seen from Fig.~\ref{fig-concentration}, this approximation matches the actual spectrum very well.

\subsubsection{Random Product States: Effects of finite $t$ and $p$}\label{sec-finitetp}

In the above, we have used \refeq{eq-approx} to approximate $\frac{1}{p}M^TM$. This approximation is accurate as $p\to\infty$ due to the law of large numbers. 

To discuss the accuracy when $p$ is finite, we need to consider the covariance of $X_\alpha X_\beta$:
\begin{equation}
    cov(X_\alpha X_\beta, X_\gamma X_\delta)=\mathbb{E}[X_\alpha X_\beta X_\gamma X_\delta]-\mathbb{E}[X_\alpha X_\beta]\mathbb{E}[X_\gamma X_\delta].
\end{equation}
Obviously $|X_\alpha|\leqslant 2$, since $\norm{\mc{O}_\alpha-\mc{O}_\alpha(t)}\leqslant 2$. Therefore $cov(X_\alpha X_\beta, X_\gamma X_\delta)=O(1)$. According to the central limit theorem, 
\begin{equation}
    \left|\frac{1}{p}(M^TM)_{\alpha\beta}-\mathbb{E}[(X^T X)]_{\alpha\beta}\right|\leqslant \frac{O(1)}{\sqrt{p}} ~(\forall\alpha, \beta),
\end{equation}
for typical sample. The maximum difference between the gap of $\frac{1}{p}(M^TM)$ ($p$ finite) and the gap of $\mathbb{E}[(X^T X)]$ ($p$ infinite) will be $\frac{O(n)}{\sqrt{p}}$. Therefore, as long as $p=\Omega(n^2)$, the approximation \refeq{eq-approx} will be reasonable.

This is just a rough (but rigorous) estimation. If $\frac{1}{p}M^TM-\mathbb{E}[(X^T X)]$ behaves like a random matrix, then typical difference of the gaps will be controlled by $O(\frac{\sqrt{n}}{\sqrt{p}})$ and $p=\Omega(n)$ will be enough to make the approximation \refeq{eq-approx} valid.  

\begin{figure}[t]
    \centering
    \begin{minipage}{0.45\textwidth}
        \centering
        \includegraphics[width=\textwidth]{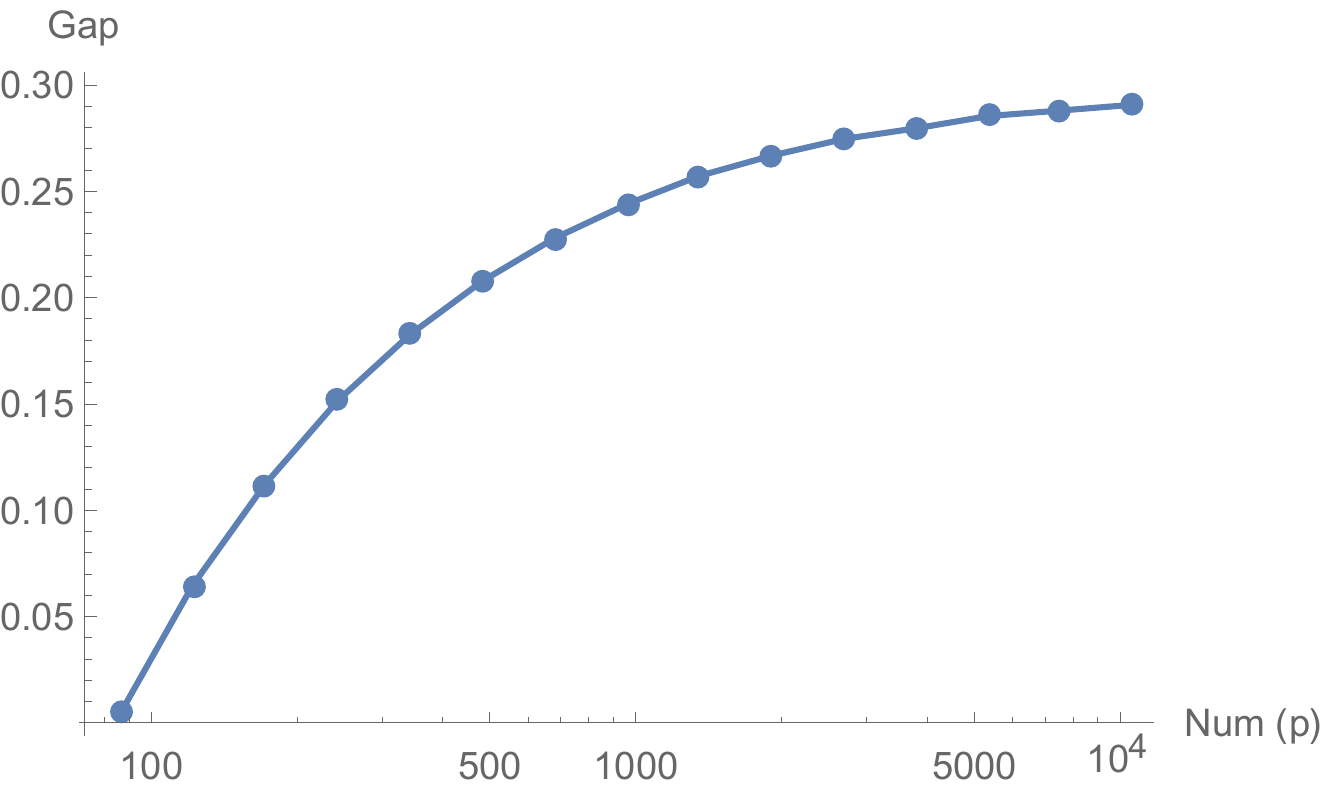}
        \caption{Dependence of gap on the number of quenches $p$. Here $L=8$, $n=87$, $t=10$.}    \label{fig-gap_p}
    \end{minipage}\hfill
    \begin{minipage}{0.45\textwidth}
        \centering
        \includegraphics[width=\textwidth]{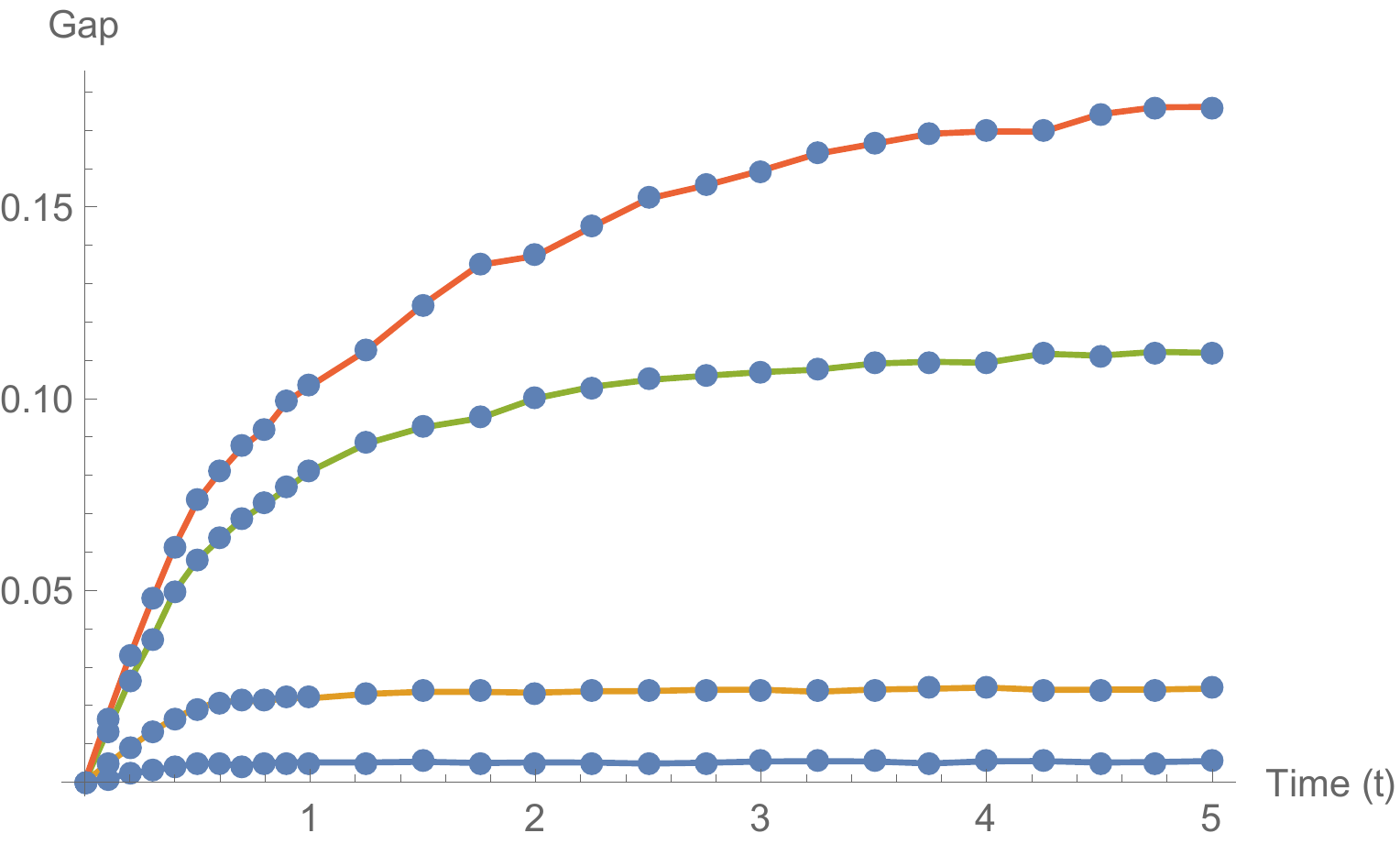}
        \caption{Dependence of gap on time. Here $L=8$. Four curves from bottom to top correspond to $p=87=n$, $p=96\approx 1.1n$, $p=174=2n$, $p=348=4n$.}\label{fig-gap_t}
    \end{minipage}
\end{figure}

In Fig.~\ref{fig-gap_p}, we plot the gap vs. $p$, when $L=8$ and $t=10$. We find that the gap grows with $p$. This is consistent with our observation that at small $p$, the error decreases faster than $\frac{1}{\sqrt{p}}$. We can also see that the gap at large $p$ is consistent with our prediction \refeq{eq-gaplowerbound}. 

In Fig.~\ref{fig-gap_t}, we plot the gap vs. $t$ for different $p$. We see that the gap grows with $t$,  consistent with the intuition that a large time interval makes the fidelity better.

\subsection{Stability against ignorance}

Assuming the exact Hamiltonian is:
\begin{equation}\label{eq-realH}
    H=\sum_{\alpha}  c_\alpha  \mc{O}_\alpha +\sum_\beta {c'}_\beta \mc{O}_\beta,
\end{equation}
but one may only pick up part of the local operators $\{\mc{O}_\alpha \}$ and obtain the following reconstructed Hamiltonian:
\begin{equation}
    H=\sum_i x_{\alpha} \mc{O}_\alpha.
\end{equation}
The reconstruction matrix now splits into two parts:
\begin{equation}
    M_{tot}=(M;M').
\end{equation}
The actual coefficient vector $\tvect{c}{c'}$ is the singular vector of $M_{tot}$ with singular value 0. By restricting the operator set to $\{\mc{O}_\alpha\}$, one only measures $M$ and 
calculates its minimal singular vector $x$. 

\begin{theorem}
The reconstruction error $\sin\theta$ between $x$ and $c$ is controlled by
\begin{equation}
    |\sin\theta|\leqslant \frac{\norm{\sqrt{\frac{1}{p}}M'{c'}}}{ s_2(\sqrt{\frac{1}{p}}M) \norm{c}}\leqslant
    \frac{\norm{\frac{1}{\sqrt{p}}M'}}{s_2\left(\sqrt{\frac{1}{p}}M\right)}
    \frac{\norm{c'}}{\norm{c}},
\end{equation}
where $s_2(\sqrt{\frac{1}{p}}M)$ denotes the second smallest singular value of $\frac{1}{\sqrt{p}}M$.
\end{theorem}

\begin{proof}
We will use $s_2$ to represent $s_2\left(\sqrt{\frac{1}{p}M}\right)$ for notational simplicity. Assuming $ s_2 $ is nonzero for now, otherwise the above inequality is trivially valid. Because $\tvect{c}{c'}$ is the coupling vector of the full Hamiltonian, $\frac{1}{\sqrt{p}}M c+\frac{1}{\sqrt{p}}M'{c'}=0$ and we have:
\begin{equation}
    \norm{\frac{1}{\sqrt{p}}Mc}=\norm{\frac{1}{\sqrt{p}}M'{c'}}\defeq\epsilon\norm{c}.
\end{equation}
 This implies $\frac{1}{\sqrt{p}}M$ has at least one singular value in $[0,\epsilon]$. Let us assume $\epsilon< s_2 $, otherwise the theorem is again trivially valid. Therefore, $\frac{1}{\sqrt{p}}M$ must have exactly one singular value in $[0,\epsilon]$ and no singular values in $(\epsilon, s_2 )$. Denote the singular vector to be $\bar{c}$ (so this is what one will get by our reconstruction method) and decompose $c$ as $c^\para+c^\perp$ with respected to $\bar{c}$. Then
\begin{equation}
    \epsilon\norm{c}=\norm{\frac{1}{\sqrt{p}}Mc}\geqslant  s_2 \norm{c^\perp}.
\end{equation}
Denote the angle between $c$ and $\bar{c}$ to be $\theta$, then
\begin{equation}
    \sin\theta=\frac{\norm{c^\perp}}{\norm{c}}
    \leqslant \frac{\epsilon}{ s_2 }=\frac{\norm{\sqrt{\frac{1}{p}}M'{c'}}}{ s_2 \norm{c}}\leqslant
    \frac{\norm{\sqrt{\frac{1}{p}}M'}}{ s_2 }
    \frac{\norm{c'}}{\norm{c}}.
\end{equation}
This is the desired upper bound for the reconstruction error.
\end{proof}

In the above expression, $\norm{\sqrt{\frac{1}{p}}M'}$ in bounded by an $O(1)$ number. Indeed, according to \refeq{eq-ACB} and \refeq{eq-CCI}, we have:
\beqa
    x^T(\frac{1}{p}M_{tot}^TM_{tot})x
    &=x^T(I-C^T)B(I-C)x+x^T\tilde{C}^T B \tilde{C}x\\
    &\leqslant \frac{1}{3}x^T\left[(I-C^T)(1-C)+\tilde{C}^T\tilde{C}\right]x\\
    &=\frac{1}{3}x^T\left(2-C^T-C\right)x.
\eeqa
According to definition \refeq{eq-operatorexpansion}, we have:
\begin{equation}
    |x^TCx|=\frac{1}{D}|\Tr\left[Q(0)^\dagger Q(t)\right]|\leqslant \frac{1}{D}\sqrt{\Tr(Q(0)^\dagger Q(0))\Tr(Q(t)^\dagger Q(t))}=\norm{x}^2.
\end{equation}
where $Q(t)=\sum x_\alpha \mc{O}_\alpha(t)$ (recall that $\Tr(\mc{O}_\alpha(0)\mc{O}_\beta(0))=D\delta_{\alpha\beta}$). Similarly $|x^TC^Tx|\leqslant \norm{x}^2$. Therefore, $x^T(\frac{1}{p}M_{tot}^TM_{tot})x\leqslant\frac{4}{3}\norm{x}^2$, which implies
\begin{equation}
    \norm{\frac{1}{\sqrt{p}}M'}\leqslant \norm{\frac{1}{\sqrt{p}}M_{tot}}\leqslant\frac{2}{\sqrt{3}}.
\end{equation}

About the second smallest singular value $s_2$, we can prove $ s_2\neq 0 $ just from the existence of a gap of $M_{tot}$.
\begin{claim} If $M_{tot}$ has a gap $\lambda$ between its minimal singular value 0 and 2nd minimal singular value, then $M$ has at most one singular value in $[0,\frac{\lambda}{\sqrt{2}})$. Therefore $ s_2 \geqslant \frac{\lambda}{\sqrt{2}}$.
\end{claim}
\begin{proof}
If not, assuming $x_1$ and $y_1$ are two (orthogonal) singular vectors of $M$ with singular values $s_x$ and $s_y$, $s_x,s_y\in[0,\frac{\lambda}{\sqrt{2}})$. Denote $x=\tvect{x_1}{0}$, $y=\tvect{y_1}{0}$. Then
\beqa
    \norm{M_{tot} x}&=\norm{M x_1}=s_x\norm{x}.
\eeqa
We decompose $x$ as $x^\para+x^\perp$ (similarly for $y$), where $\parallel$ means parallel to $\tvect{c}{c'}$, the singular vector of $M_{tot}$ with singular value 0. We have
\begin{equation}
   \frac{\lambda}{\sqrt{2}}\norm{x}> s_x\norm{x}=\norm{M_{tot}x}\geqslant \lambda \norm{x^\perp},
\end{equation}
hence $\norm{x^\perp}<\frac{1}{\sqrt{2}}\norm{x}$ and $\norm{x^\para}>\frac{1}{\sqrt{2}}\norm{x}$. Similar inequalities hold for $y$. Therefore
\begin{equation}
    |(x,y)|=|(x^\para,y^\para)+(x^\perp,y^\perp)|\geqslant|(x^\para,y^\para)|-|(x^\perp,y^\perp)|>0.
\end{equation}
However $(x,y)=(x_1,y_1)=0$, which leads to a contradiction.
\end{proof}

The gap $\lambda$ of $M_{tot}$ can be very small due to the presence of longer-range interactions in $\{\mc{O}'\}$. However, we can use the operator expansion as before to get a more reasonable estimation on $ s_2 $. In this case, \refeq{eq-MMexpansion} is still true for the $M$ here ($s\in\{\mc{O}\}$ only includes those in $\{\mc{O}\}$, not $\{\mc{O'}\}$). We also have:
\begin{equation}
     s_2 ^2\geqslant\sigma_{2}(A)=\sigma_{2}\left( (I-C^T)B(I-C)\right)\geqslant (\frac{1}{3})^{l_{\max}} \sigma_{2}\left( (I-C^T)(I-C)\right).
\end{equation}
Note that here $Cc\neq c$ so we may no longer have 0 singular value. Here $l_{\max}$ is the max length of an operator in $\{\mc{O}\}$ (instead of $\{\mc{O}\}\cup\{\mc{O}'\}$), so it is still small.

From ETH (\refeq{eq-finalvalue}), we have $C_{\alpha\beta}=\frac{ c_\alpha  c_{\beta}  }{\norm{c}^2+\norm{c'}^2}+O(\frac{1}{L^2})\defeq (C_0)_{\alpha\beta}+O(\frac{1}{L^2})$. Since the spectrum of $(I-C_0^T)(I-C_0)$ is
\begin{equation}
   \Big(1-\frac{\norm{c}^2}{\norm{c}^2+\norm{c'}^2}\Big)^2,1,1,\cdots, 1,
\end{equation}
we get:
\begin{equation}
     s_2 ^2\geqslant(\frac{1}{3})^{l_{\max}}-O(\frac{1}{L}).
\end{equation}

\bibliography{supplemental.bib}